\documentclass[twocolumn,aps,superscriptaddress,multicol,amsmath,amssymb]{revtex4-1}
\usepackage{amssymb}
\usepackage{amsmath}
\usepackage{graphicx}
\usepackage{sidecap}
\usepackage{epstopdf}
\usepackage{bm}%
\usepackage{gensymb}
\usepackage{soul}
\usepackage{sidecap}
\usepackage{bbm}
\usepackage[normalem]{ulem}

\newcommand{\bra}[1]{\ensuremath{\langle #1 |}}
\newcommand{\ket}[1]{\ensuremath{| #1 \rangle}}

\newcommand{\be}{\begin{equation}}
\newcommand{\ee}{\end{equation}}
\newcommand{\bk}{{{\bf{k}}}}
\newcommand{\bq}{{{\bf{q}}}}

\newcommand{\bg}{{{\bf{g}}}}

\newcommand{\bF}{{{\bf{F}}}}

\newcommand{\bM}{{{\bf{M}}}}

\newcommand{\bea}{\begin{eqnarray}}
\newcommand{\eea}{\end{eqnarray}}
\newcommand{\ra}{\rangle}

\newcommand{\bd}{\begin{displaymath}}
\newcommand{\ed}{\end{displaymath}}
\newcommand{\ba}{\begin{array}}
\newcommand{\ea}{\end{array}}
\newcommand{\bi}{\begin{itemize}}
\newcommand{\ei}{\end{itemize}}
\newcommand{\bc}{\begin{center}}
\newcommand{\ec}{\end{center}}
\newcommand{\bfl}{\begin{flushleft}}
\newcommand{\efl}{\end{flushleft}}
\newcommand{\bfr}{\begin{flushright}}
\newcommand{\efr}{\end{flushright}}
\newcommand{\no}{\nonumber}

\newcommand{\mi}{\rm i}
\newcommand{\DsH}{Dresselhaus}
\newcommand\redsout{\bgroup\markoverwith{\textcolor{red}{\rule[0.5ex]{2pt}{0.4pt}}}\ULon}
\newcommand\bluesout{\bgroup\markoverwith{\textcolor{blue}{\rule[0.5ex]{2pt}{0.4pt}}}\ULon}

\newcommand{\bl}{\begin{aligned}}
\newcommand{\el}{\end{aligned}}

\def\ket#1{\left\vert #1 \right\rangle}

\def\bk{{\bf k}} \def\bq{{\bf q}}  
\def\bg{{\bf g}}  \def\bd{{\bf d}}

\def\ra{\rightarrow}
\def\6{\partial}

\def\o{\omega}

\def\bra{\langle}
\def\ket{\rangle}
\def\={\!\!\!&=&\!\!\!}
\def\+{\!\!\!&&\!\!\!+~}
\def\-{\!\!\!&&\!\!\!-~}

 \newcommand\redout{\bgroup\markoverwith{\textcolor{red}{\rule[.5ex]{2pt}{0.4pt}}}\ULon}

\usepackage[colorlinks=true,citecolor=blue]{hyperref}
\hypersetup{colorlinks=true,citecolor=blue,linkcolor=red,urlcolor=blue}

\usepackage{hyperref,cleveref}
\begin{document}
\title{
 Superconductivity of mixed parity and frequency  in an anisotropic spin-orbit coupling}

\author{Mehdi Biderang} 
\email{mehdi.biderang@apctp.org}
\affiliation{Asia Pacific Center for Theoretical Physics, Pohang, Gyeongbuk 790-784, Korea}

\author{Mohammad-Hossein Zare}
\affiliation{Department of Physics, Faculty of Science, Qom University of Technology, Qom 37181-46645, Iran}

\author{Alireza Akbari}
 \email{alireza@apctp.org}
 \affiliation{Asia Pacific Center for Theoretical Physics, Pohang, Gyeongbuk 790-784, Korea}
 \affiliation{Department of Physics, POSTECH, Pohang, Gyeongbuk 790-784, Korea}
\affiliation{Max Planck POSTECH Center for Complex Phase Materials, POSTECH, Pohang 790-784, Korea}

\date{\today}
%
\begin{abstract}
We  illuminate the superconducting phases in [001]-grown-noncentrosymmetric quantum wells with an anisotropic spin-orbit coupling in the presence of on-site Hubbard interaction.
Within the random phase approximation, we investigate  the spin-fluctuation-mediated pairing 
 in the presence of Rashba/\DsH~ antisymmetric spin-orbit couplings. %
Although the existence of spatial inversion symmetry  desires a dominant $d$-wave pairing for all filling levels,
 a broken inversion symmetry generates antisymmetric spin-orbit coupling and mixes the even- and odd-parity in the superconducting gap. %
We study the symmetry of the mixed-parity gap for various strengths of Hubbard interaction.
Besides, we consider a superconductor-ferromagnet junction to survey the modifications of superconducting order parameters, and  observe  %
an admixture of even- and odd-frequencies   due to the ferromagnet exchange field.
\end{abstract}
%
\maketitle
%
\section{Introduction}
The superconducting phase of interacting electrons has attracted high attention by discovery of unconventional superconductors,  which after more than three decades, still  
 the origin and nature of the pairing symmetry  are  unclear~\cite{Sigrist_AIP_2005,Schriffer_PRB_1989,Anderson_Phys_Rev_1960,Maiti_Chubukov_2013}.
In most cases, the
magnetism and  superconductivity   are coexisting/competing  via strong magnetic fluctuations~\cite{Sigrist_Rev_Mod_Phys_1991,Bauer_Sigrist_NCS_Book_2012,Bauer_Sigrist_PRL_2004}, and 
the  electron-phonon mechanism is not  gluing the Cooper pairs~\cite{Mineev_1999,Scalapino_Rv_Mod_Phys_2012}.
This interplay of magnetic fluctuations and superconductivity proposes a  formation of Cooper pairs with higher angular momentum like p-wave pairing in ${}^3$He or d-wave in high-temperature superconductors~\cite{Annett_1995,Scalapino_Low_Temp_Phys_1999}.

Understanding of the superconductivity has undergone a significant progress by studying the  instabilities driven by repulsive electron-electron interactions, especially in the weak coupling limit~\cite{Scalapino_PRB_1986,Greco_Schnyder_PRL_2018,Ghadimi_PRB_2019}.
 Whereas this short-ranged  repulsive interaction may achieve an anisotropic pairings near the antiferromagnetic/spin-density-wave  
 instability~\cite{Hirsch_PRL_1985317,Romer_Eremin_PRB_2015}.
A powerful  tool  to investigate them  is the symmetry analysis, which remains reliable even in the ambiguity of the pairing mechanism.
Particularly, the discrete symmetries, such as time-reversal symmetry (TRS) or inversion symmetry (IS)  play important roles~\cite{Rossi_PRL_2016,Annica_PRB_2018,Linder_arXiv_2019,Zare_PRB_2018}  
and would affect the properties of superconducting gap function~\cite{Bauer_Sigrist_PRL_2004,Brydon_PRB_2011,Goryo_2014,Brydon_PRB_2012,Biderang_PRB_2018,Annett_2018}.
In that matter, if at least one of these symmetries  is broken, Cooper pairs, which have an antisymmetric wave functions under the interchange of all quantum numbers,
acquire exotic forms.
As an example, in ferromagnetic superconductors with broken TRS, the electrons are most likely paired in the same spin directions~\cite{Bauer_Sigrist_NCS_Book_2012}. 
Thus lack of TRS may cause to sign changing of the Cooper pair's wave function under time-exchange.
This unequal time pairing yields odd-frequency superconducting couples  
for describing superfluidity~\cite{Berezinskii_1974}, as well as, 
 superconductivity~\cite{Balatsky_PRB_1992}.
Moreover, there are many proposals for the possibility of superconductor-ferromagnet (FM) junctions to realize induced odd-frequency pairing, or  the admixture of even- and odd- frequency Cooper pairings~\cite{Linder_Nat_Phys_2015,Jeon_Nat_Mat_2018,Zare_PRB_2018}.

Recently, topological superconductivity in heterostructures with strong  spin-orbit coupling (SOC) have found a wide range of promising applications~\cite{Brydon_PRL_2013,Brydon_NJP_2013,Scarf_PRB_2019}.
In noncentrosymmetric superconductors (NCSs) with violated parity, the lack of IS leads to mixing of even- and odd-parity pairings~\cite{Samokhin_Mineev_PRB_2008,Sato_PRB_2009,Yanase_PRB_2018,Yanase_PRB_2018_2}.
Absence of inversion center in these materials induces an antisymmetric SOC, which has interesting consequences on the electronic structure of the system~\cite{Samokhin_PRB_2003,Bauer_Sigrist_PRL_2004,Frigeri_Sigrist_PRL_2004}.
Considering the origin, there are two  types of antisymmetric SOCs in noncentrosymmetric systems, which are derived from a bulk (\DsH)~\cite{Dresselhaus_Phys_Rev_1955} or a structure (Rashba)~\cite{Rashba_Sov_Phys_1960} inversion asymmetry.
The effects of the Rashba/\DsH~SOCs have been reviewd in a variety of systems like ultracold fermions~\cite{Hu_PRL_2011,ZQ_Yu_PRL_2011,Dell_Anna_PRA_2012,Z_Li_PRB_2012,L_He_PRL_2012}, quantum wells~\cite{Ganichev_PRL_2004,Koralek_Nature_2009}, two-dimensional (2D) NSC systems~\cite{Dias_2016}, Weyl semimetals~\cite{Biderang_Drumhead_PRB_2018}, transition metal dichalcogenides~\cite{Xiao_PRL_2012}, and half-Heusler compounds~\cite{Tafti_PRB_2013}.

 More exciting, the cooperative  Rashba and \DsH~\cite{Engels_PRB_1997,Nitta_PRL_1997} can come up with a  possibility of topological superconductivity and topological edge states~\cite{Biderang_PRB_2018,Ikegaya:2017aa}.
An exotic platform for the recognition, is in ultracold superfluid Fermi gases, where a synthetic mixture antisymmetric SOCs
 may be created and tuned by the inversion symmetry breaking caused by an applied laser field~\cite{Sau_PRB_2011}.
 Another possible case may realize by applying an external bias voltage or using the monolayer in heterostructures may lead to an induced Rashba-type 
along with the \DsH,  in transition metal dichalcogenides~\cite{Xiao_PRL_2012}.
 Lifting of spin degeneracy consideres these ultrathin materials  as potential candidates to realize topological superconductivity, and to effectively host spinless fermions as a result of the broken inversion symmetry~\cite{Vaezi_Nature_2017}.
Another promising examples are half-Heusler compounds, where the zinc-blend structure induces an antisymmetric \DsH. 
Their superconducting order parameter has an even-odd parity mixing, %
and they can fabricate a  peculiar spin texture of topological surface states~\cite{Scarf_PRB_2019,Brydon_PRL_2016,Butch_PRB_2011}.  
\\

In this paper, we consider the on-site Hubbard model into a 2D
metallic
 noncentrosymmetric square lattice with coincidence of Rashba or/and \DsH~ 
without any charge or magnetic order.
Notably, we aim to inspect the role of antisymmetric SOCs and electron filling on the symmetry of superconducting gap function at different values of correlation.
We survey the spin-fluctuation-mediated pairing interaction within random phase approximation (RPA) to form mixed singlet-triplet superconductivity.
Moreover, we examine the effect of an NCS-FM heterostructure on the symmetry of Cooper pairs and verify the emergence of mixed even- and odd-frequency superconductivity in the NCS due to the induced broken of TRS.
%

\begin{figure*}[t]
	\begin{center}
		\hspace{0.1cm}
		\includegraphics[width=0.99 \linewidth]{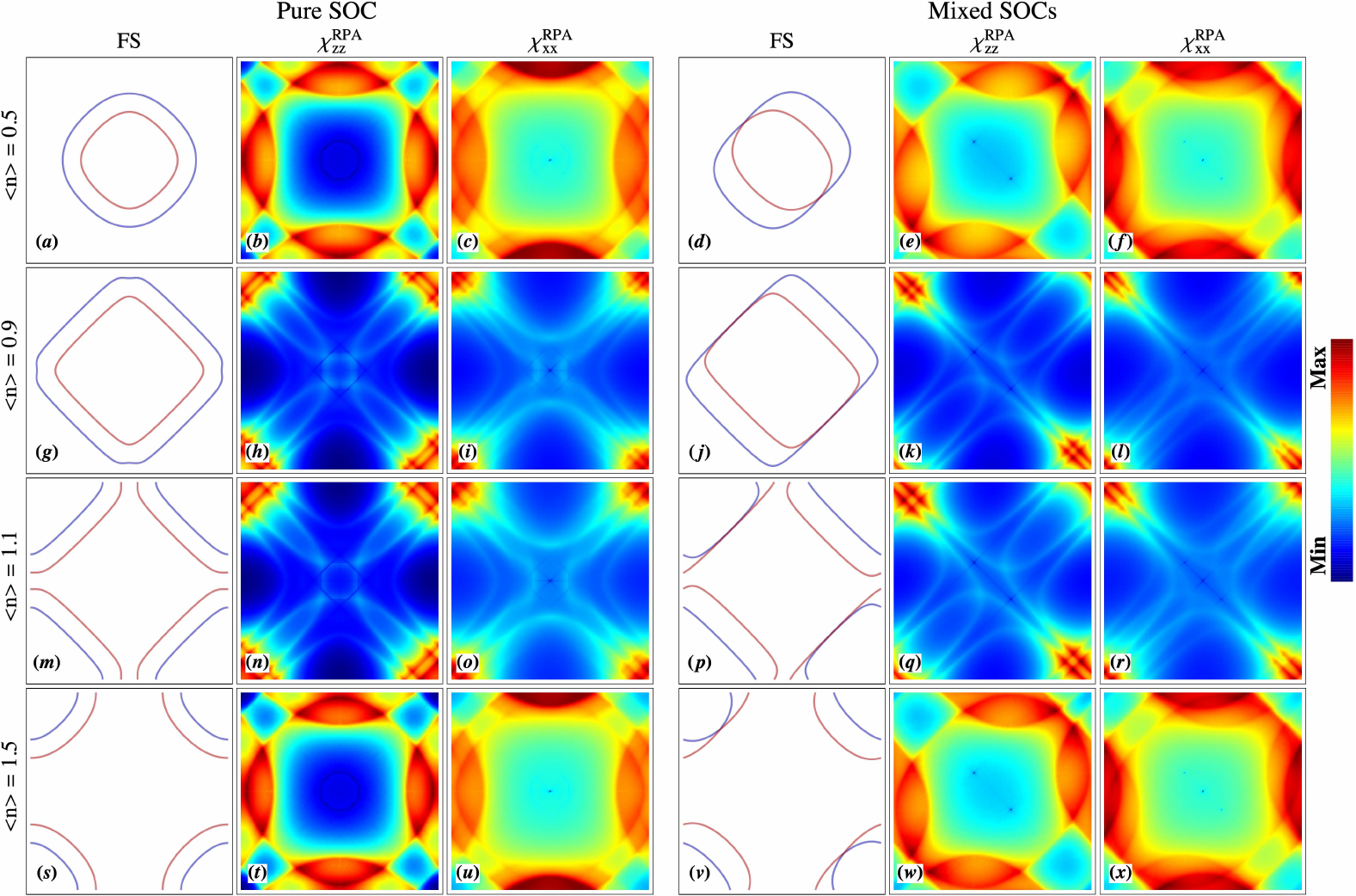} 
	\end{center}
		\vspace{-0.3cm}
\caption{(Color online) 
		Fermi surfaces, longitudinal and transverse components of RPA susceptibility for $\omega=0$ for different values of filling $\langle n \rangle$ in the presence of antisymmetric SOCs.
		Left panel: for a square system with pure Rashba (\DsH), and
		right panel: with the same amplitudes of Rashba and \DsH.
		The red (blue) line indicates the positive (negative), $\varepsilon^{}_{\bk+}=0$ ($\varepsilon^{}_{\bk-}=0$),  helical Fermi surfaces.
		The FS has anisotropic structure within the BZ for coincidence of Rashba and \DsH. 
		The middle and right columns at each panel represent the longitudinal ($\chi^{\rm RPA}_{zz}(\bq,\omega=0)$) and transverse ($\chi^{\rm RPA}_{xx}(\bq,\omega=0)$) components of RPA spin susceptibility for $U=0.5t$, respectively.
		The in-plane antisymmetric SOC causes an anisotropy in spatial components of spin susceptibility, i.e. $\chi^{\rm RPA}_{xx} \neq \chi^{\rm RPA}_{yy} \neq \chi^{\rm RPA}_{zz}$~\cite{Cobo_PRB_2016}.
		It should be mentioned that $\chi^{\rm RPA}_{yy}$ is obtained from rotation of $\chi^{\rm RPA}_{xx}$ with a counter-clockwise $90^{\circ}$ angle.
		Note: The axes in FS and susceptibility plots are $(k_x, k_y)$ and $(q_x, q_y)$, both in the $[-\pi,\pi]$ interval.
	}
	\label{Fig:FS}
\end{figure*}
%

%
%
\begin{figure}[t]
	\begin{center}
		\vspace{0.cm}
		\hspace{-0.2cm}
		\includegraphics[width=0.99 \linewidth]{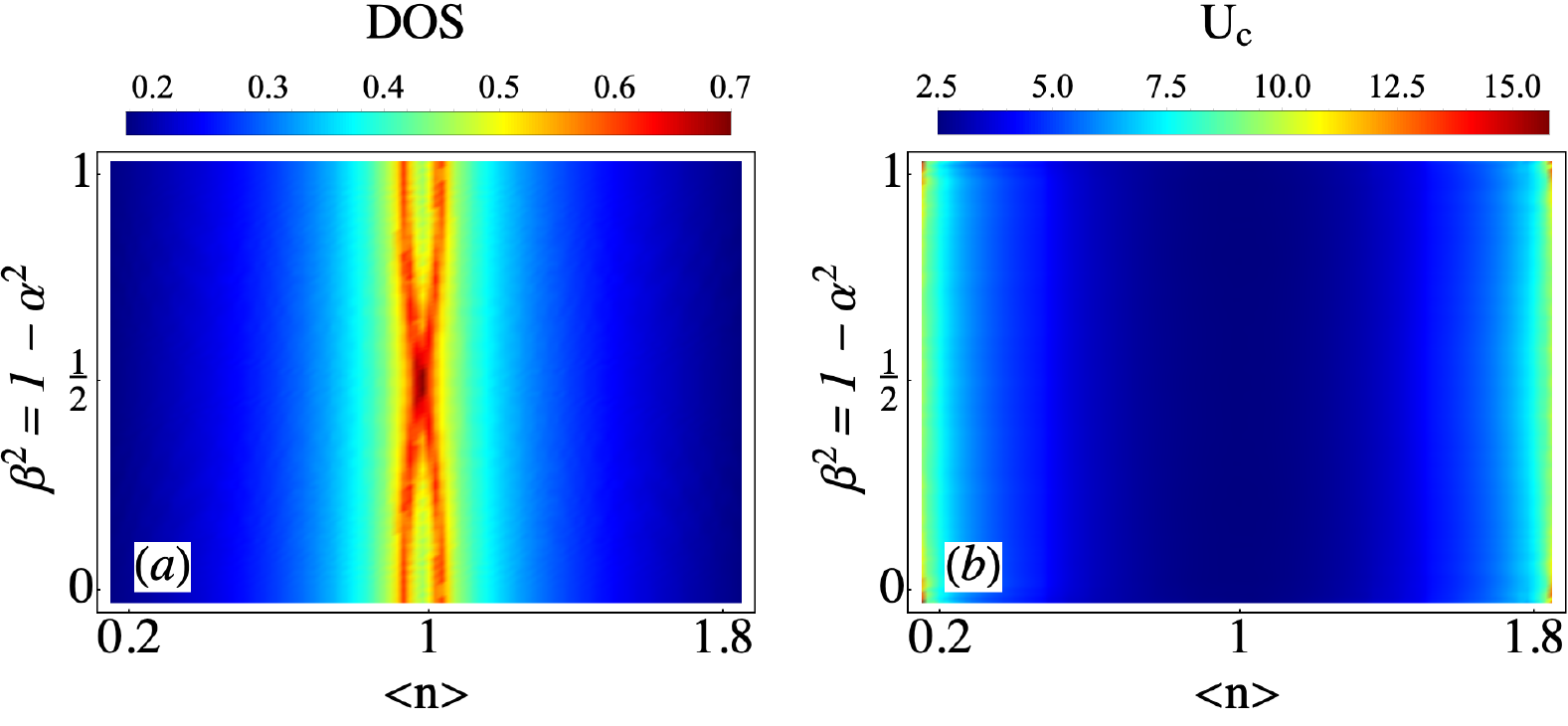} 
	\end{center}
	\vspace{-0.3cm}
	\caption{(Color online) 
		(a) Density of states (DOS) at the Fermi surface ($\omega=0$).
		 (b) The values of critical interaction as a function of filling $\bra n\ket$ and relative strengths of Rashba and \DsH. 
		It is obviously seen that the van Hove singularities occur near the half-filling.
	}
	\label{Fig:DOS_Uc}
\end{figure}
%

\section{Theoretical Model  Hamiltonian}
We consider a 2D single band noncentrosymmetric system on the square lattice in the presence of on-site Hubbard interaction.
The Hamiltonian is given by
\
\begin{align}
\begin{aligned}
{\cal H}^{}_{0}=
&\sum_{\bk,ss'}
\Big[
\epsilon_{\bk}\sigma^{}_0+{\bg}^{}_{\bk}\cdot {\bm \sigma}
\Big]^{}_{ss'}
c^{\dagger}_{\bk s} c^{}_{\bk s'}
\\
&
+
\frac{U}{2}\sum_{\bk\bk'\bq,s}
c^{\dagger}_{\bk s}c^{}_{\bk+\bq s}
c^{\dagger}_{\bk' \bar{s}}c^{}_{\bk'-\bq \bar{s}},
\end{aligned}
\label{Eq:Ham}
\end{align}
where $c^{\dagger}_{\bk s}$ ($c^{}_{\bk s}$) creates (annihilates) an electron with momentum $\bk$ and spin $s=\uparrow,\downarrow$. Here   $\bar{s}$ defines spin with the opposite direction of $s$, and also we set the lattice parameter  to $1$.
Moreover, $\sigma^{}_0$ and $\sigma^{}_i$ ($i=x,y,z$) represent the unitary and Pauli's matrices in the spin basis, respectively.  
The Hamiltonian includes a noninteracting and Hubbard parts, which the former consist of a kinetic term $\epsilon^{}_{\bk}=-\mu-2t(\cos k_x+\cos k_y)$ and an antisymmetric SOC, and the latter denotes the on-site Hubbard interaction with strength $U$.

In quasi-two-dimensional systems, due to the dimensional confinement and symmetry reduction,
the leading term for a SOC is linear in momentum~\cite{SCHLIEMANN_2006,Ganichev_2014}.
Therefore, the SOC $\bg$-vector is antisymmetric with respect to the momentum  ($\bg^{}_{-\bk}=-\bg^{}_{\bk}$) and characterised by 
%
\be 
\no
{\bg}^{}_{\bk}=\alpha {\bg^{R}_{\bk}}+\beta  {\bg^D_{\bk}},
\ee
where the corresponding Rashba and \DsH~ 
  $\bg$-vectors are defined  by 
\bea
\no
{\bg^{R}_{\bk}}={\rm g}(\sin{k_y},-\sin{k_x});
\\\no
{\bg^D_{\bk}}={\rm g}(\sin{k_x},-\sin{k_y}).
\eea
 %
 Here, ${\rm g}$ denotes the magnitude of SOC and  the normalized  control parameters   
  $ |\alpha|, |\beta| \in [0,1]$;  ($\alpha^2+\beta^2=1$), 
 tune  the magnitudes of Rashba and \DsH. 
 \\

 The spin and orbital degrees of freedom are entangled together by antisymmetric SOC.
 Then, the two-fold spin degeneracy is lift and Fermi surface (FS) splits into two opposite-helicity ($\xi=\pm$) sheets.
The energy spectrum of each helical FS is given by
%
%
\begin{align}
\begin{aligned}
\varepsilon^{}_{\bk\pm}
\!
=
\!
\epsilon^{}_{\bk} 
\pm
{\rm g} \sqrt{
\sin^2{k_x}+\sin^2{k_y} +4\alpha\beta \sin{k_x} \sin{k_y}}~.
\end{aligned}
\label{Eq:Normal_dispersion}
\end{align}
%
We show the evolution of Fermi surface at the particular levels of filling  for the pure Rashba (\DsH)  and mixed SOCs  in first and fourth columns of   Fig.~\ref{Fig:FS},  respectively.
For the numerical calculations, we set  ${\rm g}=0.6t$.
 As it is seen, in the case of  pure Rashba or pure \DsH, the Fermi surface is isotropic structure entire the Brillouin zone (BZ) and it has a  $C_{4v}$ point group symmetry.
However, the mixing of Rashba and \DsH~ causes to reduction of symmetry group to $C_{2v}$ with an anisotropic structure of Fermi surfaces.
For cases with $\alpha=\pm \beta$, the helical bands touch each other at certain directions $[1 \bar{1}]$ and $[1 1]$.

Fig.~\ref{Fig:DOS_Uc}(a) displays density of states (DOS) at Fermi surface ($\omega=0$) at different levels of filling $\bra n \ket$ and relative strengths of Rashba and \DsH.
For the systems with dominant Rashba (\DsH), two van Hove singularities occur for both electron- and hole-doped cases near the half-filling.
In the particular case with the same magnitudes of Rashba and \DsH, the van Hove singularity only emerges at half-filling.
%
%

\section{ Spin Susceptibility and Effective Interaction: RPA study}
In this section, we survey the effective interaction between electron using RPA approach.
The pairing interaction can be extracted from the higher order diagrams of the Coulomb repulsion~\cite{Berk_PRL_1966,Scalapino_PRB_1986,Romer_Eremin_PRB_2015}.
These terms can be expressed by spin susceptibility.
The bare spin susceptibility in spin basis is expressed as
%
%
\begin{align}
\begin{aligned}
&
\chi^0_{s^{}_1s^{}_2s^{}_3s^{}_4}
(\bq,{\mi}\omega_n)=
\\
&
\;\;\;
-\frac{T}{N}\sum_{\bk,{\mi}\nu_m}
{\cal G}^0_{s^{}_1s^{}_2}(\bk,{\mi}\nu_m)
{\cal G}^0_{s^{}_3s^{}_4}(\bk+\bq,{\mi}\nu_m+{\mi}\omega_n),
\end{aligned}
\label{Eq:Bare_Susc}
\end{align}
%
with
the free particle Green's function in the spin basis is given by
%
%
\begin{align}
\begin{aligned}
{\cal G}^0_{ss'}(\bk,{\mi}\nu_m)=
\Big[
	(
	{\mi}\nu_m-\epsilon^{}_{\bk}
	)\sigma_0
-
{\bg}^{}_{\bk}\cdot {\bm \sigma}
\Big]^{-1}_{ss'}.
\end{aligned}
\label{Eq:Bare_Green}
\end{align}
%
Here 
 ${\mi}\omega_n=2n{\mi}\pi T$ and ${\mi}\nu_m=(2m+1){\mi}\pi T$ are imaginary bosonic and fermionic Matsubara frequencies at temperature $T$, respectively.
\\

In the framework of RPA, the dressed spin susceptibility is obtained by
%
%
\begin{equation}
\hat{\chi}^{\rm RPA}_{}(\bq,{\mi}\omega_n)=
\Big[
	\mathbbm{1}-\hat{\chi}^0_{}(\bq,{\mi}\omega_n)\hat{U}
\Big]^{-1}
\;
\hat{\chi}^0_{}(\bq,{\mi}\omega_n)
.
\label{Eq:RPA_Susc}
\end{equation}
%
Here $\mathbbm{1}$ is a unitary $4\times 4$ matrix and $\hat{\chi}^0_{}(\bq,{\mi}\omega_n)$ is the matrix including the sixteen components of the bare spin susceptibility with the following form
%
%
\bea
\begin{aligned}
\hspace{-1cm}
\hat{\chi}^{0}_{}(\bq,{\mi}\omega_n)=
\begin{bmatrix}
\chi^{0}_{\uparrow\uparrow\uparrow\uparrow}
&
\chi^{0}_{\uparrow\downarrow\uparrow\uparrow}
&
\chi^{0}_{\uparrow\uparrow\downarrow\uparrow}
&
\chi^{0}_{\uparrow\downarrow\downarrow\uparrow}
\\
\chi^{0}_{\uparrow\uparrow\uparrow\downarrow}
&
\chi^{0}_{\uparrow\downarrow\uparrow\downarrow}
&
\chi^{0}_{\uparrow\uparrow\downarrow\downarrow}
&
\chi^{0}_{\uparrow\downarrow\downarrow\downarrow}
\\
\chi^{0}_{\downarrow\uparrow\uparrow\uparrow}
&
\chi^{0}_{\downarrow\downarrow\uparrow\uparrow}
&
\chi^{0}_{\downarrow\uparrow\downarrow\uparrow}
&
\chi^{0}_{\downarrow\downarrow\downarrow\uparrow}
\\
\chi^{0}_{\downarrow\uparrow\uparrow\downarrow}
&
\chi^{0}_{\downarrow\downarrow\uparrow\downarrow}
&
\chi^{0}_{\downarrow\uparrow\downarrow\downarrow}
&
\chi^{0}_{\downarrow\downarrow\downarrow\downarrow}
\end{bmatrix},
\end{aligned}
\label{Eq:Bare_Susc_Mat}
\eea
%
in which, its individual  components are defined based on  the following  Lindhard function
%
%
\begin{align}
\begin{aligned}
&\chi^0_{s_1s_2s_3s_4}(\bq,{\mi}\omega_n)=
\\
&\hspace{1cm}
\frac{1}{4N}\sum_{\bk,\xi\xi'} 
\Big[ \zeta \Big]^{s^{}_1s^{}_2s^{}_3s^{}_4}_{\bk\bq,\xi\xi'}
\frac{
	f(\varepsilon^{}_{\bk+\bq,\xi'})
	-
	f(\varepsilon^{}_{\bk,\xi})
}
{
	{\mi}\omega_n
	+
\varepsilon^{}_{\bk,\xi}
-
\varepsilon^{}_{\bk+\bq,\xi'}
},
\end{aligned}
\label{Eq:Susc_Elements}
\end{align}
%
with
%
%
\begin{align}
\begin{aligned} 
\Big[ \zeta \Big]^{s^{}_1s^{}_2s^{}_3s^{}_4}_{\bk\bq,\xi\xi'}=
\Big[
\sigma_0+\xi \hat{\bg}^{}_{\bk}\cdot {\bm \sigma}
\Big]^{}_{s^{}_1s^{}_2}
\Big[
\sigma_0+\xi' \hat{\bg}^{}_{\bk+\bq}\cdot {\bm \sigma}
\Big]^{}_{s^{}_3s^{}_4},
\end{aligned}
\label{Eq:Coeff}
\end{align}
%
and $\hat{\bg}^{}_{\bk}=\bg^{}_{\bk}/|\bg^{}_{\bk}|$.
Consequently, the spatial components of RPA susceptibilities are expressed as
%
%
\begin{align}
\begin{aligned}
\chi^{\rm RPA}_{xx}(\bq,{\mi}\omega_n)
&=
\sum_{s}
\Big[
\chi^{\rm RPA}_{ss\bar{s}\bar{s}}(\bq,{\mi}\omega_n)
+
\chi^{\rm RPA}_{s\bar{s}s\bar{s}}(\bq,{\mi}\omega_n)
\Big],
\\
\chi^{\rm RPA}_{yy}(\bq,{\mi}\omega_n)
&=
\sum_{s}
\Big[
\chi^{\rm RPA}_{ss\bar{s}\bar{s}}(\bq,{\mi}\omega_n)
-
\chi^{\rm RPA}_{s\bar{s}s\bar{s}}(\bq,{\mi}\omega_n)
\Big],
\\
\chi^{\rm RPA}_{zz}(\bq,{\mi}\omega_n)
&=
\sum_{s}
\Big[
\chi^{\rm RPA}_{ssss}(\bq,{\mi}\omega_n)
+
\chi^{\rm RPA}_{s\bar{s}\bar{s}s}(\bq,{\mi}\omega_n)
\Big].
\end{aligned}
\label{Eq:Long_Trans_Susc}
\end{align}
%

By setting  $U=0.5t$  and performing the Matsubara frequency summation over ${\mi}\omega_n \ra \omega+ {\mi}0^+$,
in  Fig.~\ref{Fig:FS}, we present  the longitudinal, 
$\chi^{\rm RPA}_{zz}(\bq,\omega=0)$, 
and transverse, 
$\chi^{\rm RPA}_{xx}(\bq,\omega=0)$, 
components of the RPA spin susceptibility for several levels of filling $\langle n \rangle$.
The second and third (fifth and sixth) columns represents the  longitudinal, 
and transverse susceptibilities for the pure (mixed) SOCs, respectively.
%
The anisotropy in longitudinal and transverse spin susceptibilities are resulted from the breaking of spin rotation symmetry made by antisymmetric SOCs. 
For pure Rashba (\DsH), both longitudinal and transverse susceptibilities show nearly incommensurate fluctuations at high levels of filling.
However, near half-filling the susceptibility patterns exhibit roughly commensurate antiferromagnetic fluctuations.
It should be noted that the symmetry in spin susceptibility of electron- and hole-doped cases is presented because of  the intrinsic electron-hole symmetry of Eq.~(\ref{Eq:Ham}). 
Coexistence of Rashba and \DsH~ generates an anisotropy in spin fluctuations with incommensurate antiferromagnetic vector besides the emergence of the sub-dominant transverse and longitudinal fluctuations.
The critical on-site Hubbard interaction, $U_c$, as a function of filling $\bra n\ket$ and relative strengths of Rashba and \DsH~ is shown Fig.~\ref{Fig:DOS_Uc}(b), for which both longitudinal and transverse spin susceptibilities simultaneously diverge.

%
Since we only consider the weak couplings, it is sufficient to take into account the static susceptibility ($\omega=0$) and its driven pairing interaction~\cite{Greco_Schnyder_PRL_2018}.
In the case with zero spin-orbit coupling (${\rm g}=0$), the $O(3)$ symmetry is satisfied and the physical parts of spin susceptibility are the same ($\chi^{\rm RPA}_{xx}=\chi^{\rm RPA}_{yy}=\chi^{\rm RPA}_{zz}$).
In the presence of Ising-like spin-orbit coupling with $L_zS_z$ form, an anisotrpopy is induced between longitudinal and transverse susceptibilities, i.e. $\chi^{\rm RPA}_{xx}=\chi^{\rm RPA}_{yy}\neq \chi^{\rm RPA}_{zz}$.
However, Rashba and/or \DsH, ($S_{+}L_{-}+S_{-}L_{+}$), generate a full anisotropy among the spatial susceptibilities
($\chi^{\rm RPA}_{xx}\neq \chi^{\rm RPA}_{yy}\neq \chi^{\rm RPA}_{zz}$)~(for more related discussions see Ref.~\cite{Cobo_PRB_2016}).
Under this basis, the interaction matrix $\hat{U}$ finds an off-diagonal shape as
%
%
\begin{align}
\begin{aligned}
\hat{U}=
\begin{bmatrix}
0 & 0 & 0 & U
\\
0 & 0 & -U & 0
\\
0 & -U & 0 & 0
\\
U & 0 & 0 & 0
\end{bmatrix},
\end{aligned}
\label{Eq:U_Mat}
\end{align}
%
and the effective interaction within RPA is given by
%
%
\begin{equation}
\hat{U}^{}_{\rm eff}(\bq)=
\Big[
	\mathbbm{1}-\hat{\chi}^0_{}(\bq,{\mi}\omega_n)\hat{U}
\Big]^{-1}
\;
\hat{U}
.
\label{Eq:Eff_Int}
\end{equation}
%
It should be mentioned that we merely consider the contribution of spin fluctuations in calculation of the pairing attraction.
The effective interaction for opposite electron spins consist of an even number of bubbles and ladder diagrams.
However, for electrons with same spins, the effective interaction is explained as a summation over an odd number of bubble diagrams~\cite{Scalapino_PRB_1986}.
Thus, the effective spin-fluctuation-mediated interactions $\Gamma^{\rm eff}_{\rm same}(\bk,\bk')$ and $\Gamma^{\rm eff}_{\rm opp}(\bk,\bk')$ for both same and opposite spins within RPA are written as
%
%
\begin{align}
\begin{aligned}
&
\Gamma^{\rm eff}_{\rm same}(\bk,\bk')
=
\frac
{
	U^2 \Big(
	\chi^0_{\uparrow\uparrow\uparrow\uparrow}(\bk-\bk')
	+
	\chi^0_{\downarrow\downarrow\downarrow\downarrow}(\bk-\bk')
	\Big)
}{
1-U
\Big(
\chi^0_{\uparrow\downarrow\downarrow\uparrow}(\bk-\bk')
+
\chi^0_{\downarrow\uparrow\uparrow\downarrow}(\bk-\bk')
\Big)
},
\\
&
\Gamma^{\rm eff}_{\rm opp }(\bk,\bk')
=
U \Bigg[
\frac
{
	2-U
	\Big(
	\chi^0_{\uparrow\downarrow\downarrow\uparrow}(\bk-\bk')
	+
	\chi^0_{\downarrow\uparrow\uparrow\downarrow}(\bk-\bk')
	\Big)
}{
	1-U
	\Big(
	\chi^0_{\uparrow\downarrow\downarrow\uparrow}(\bk-\bk')
	+
	\chi^0_{\downarrow\uparrow\uparrow\downarrow}(\bk-\bk')
	\Big)
}
\\
&
\hspace{1.6cm}
-
\frac
{
	 2+U
	\Big(
	\chi^0_{\uparrow\uparrow\downarrow\downarrow}(\bk+\bk')
	+
	\chi^0_{\downarrow\downarrow\uparrow\uparrow}(\bk+\bk')
	\Big)
}{
	1+U
	\Big(
	\chi^0_{\uparrow\uparrow\downarrow\downarrow}(\bk+\bk')
	+
	\chi^0_{\downarrow\downarrow\uparrow\uparrow}(\bk+\bk')
	\Big)
}
	\Bigg].
\end{aligned}
\label{Eq:Effective_Interaction}
\end{align}
%
%
Note that  $\Gamma^{\rm eff}_{\rm same}$ and the first term in $\Gamma^{\rm eff}_{\rm opp}$ are related to the bubble diagram contributions.
However, the second part of $\Gamma^{\rm eff}_{\rm opp}$ corresponds to the ladder diagrams.
%

%
\begin{figure}[t]
	\begin{center}
		\vspace{0.cm}
		\hspace{-0.1cm}
		\includegraphics[width=1.04 \linewidth]{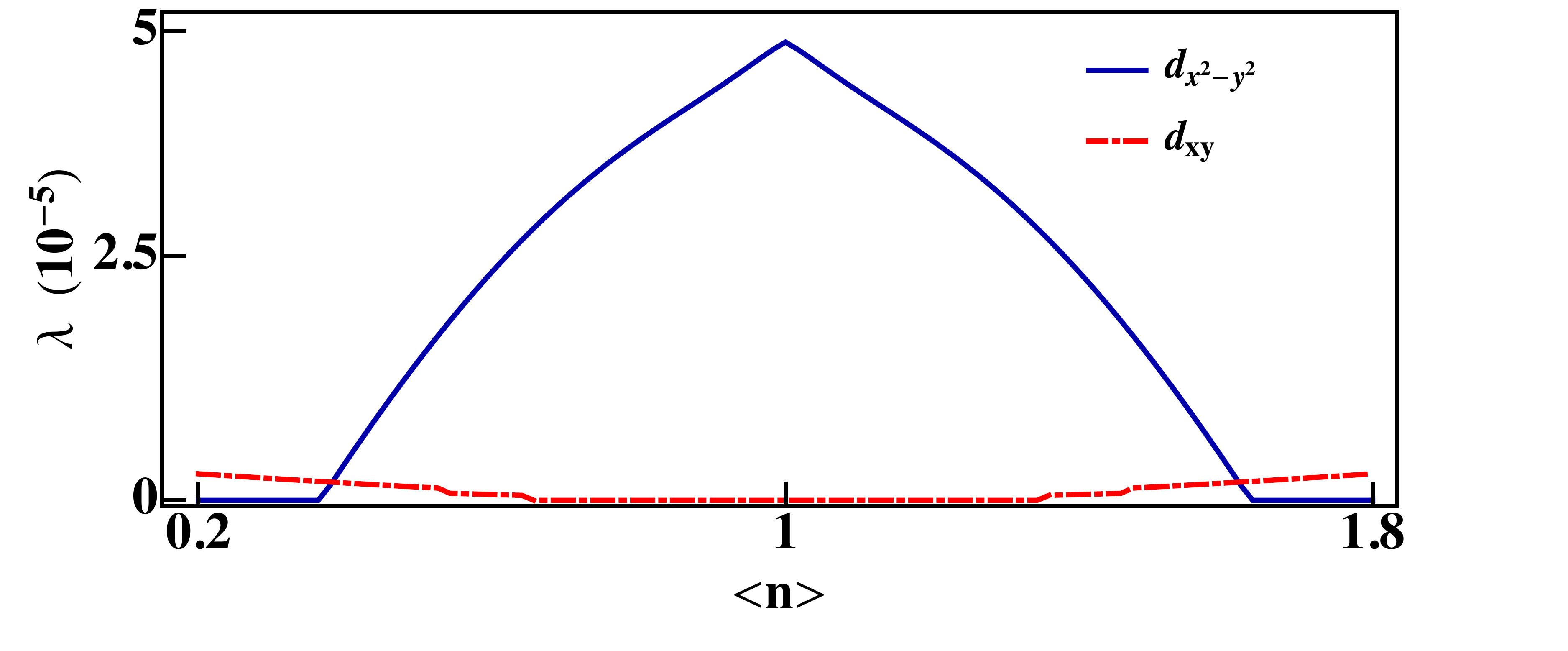} 
	\end{center}
	\vspace{-0.8cm}
	\caption{(Color online) 
		Filling dependence of the effective coupling constant $\lambda^{\rm eff}_i$ in the absence of antisymmetric SOCs (${\rm g}=0$), in the singlet channel for $U=0.5t$.
		The superconducting gap function has $d$-wave nature for entire of filling range.
		It is shown that for a wide range of filling, the $d_{x^2-y^2}$-wave pairing dominates over $d_{xy}$-wave gap symmetry.
		However, in the regions far away from half-filling, $d_{xy}$-wave pairing is the leading order.
	}
	\label{Fig:Lambda_g_0}
\end{figure}
%

%
\begin{figure*}[t]
	\begin{center}
		\vspace{0.cm}
		\includegraphics[width=0.99 \linewidth]{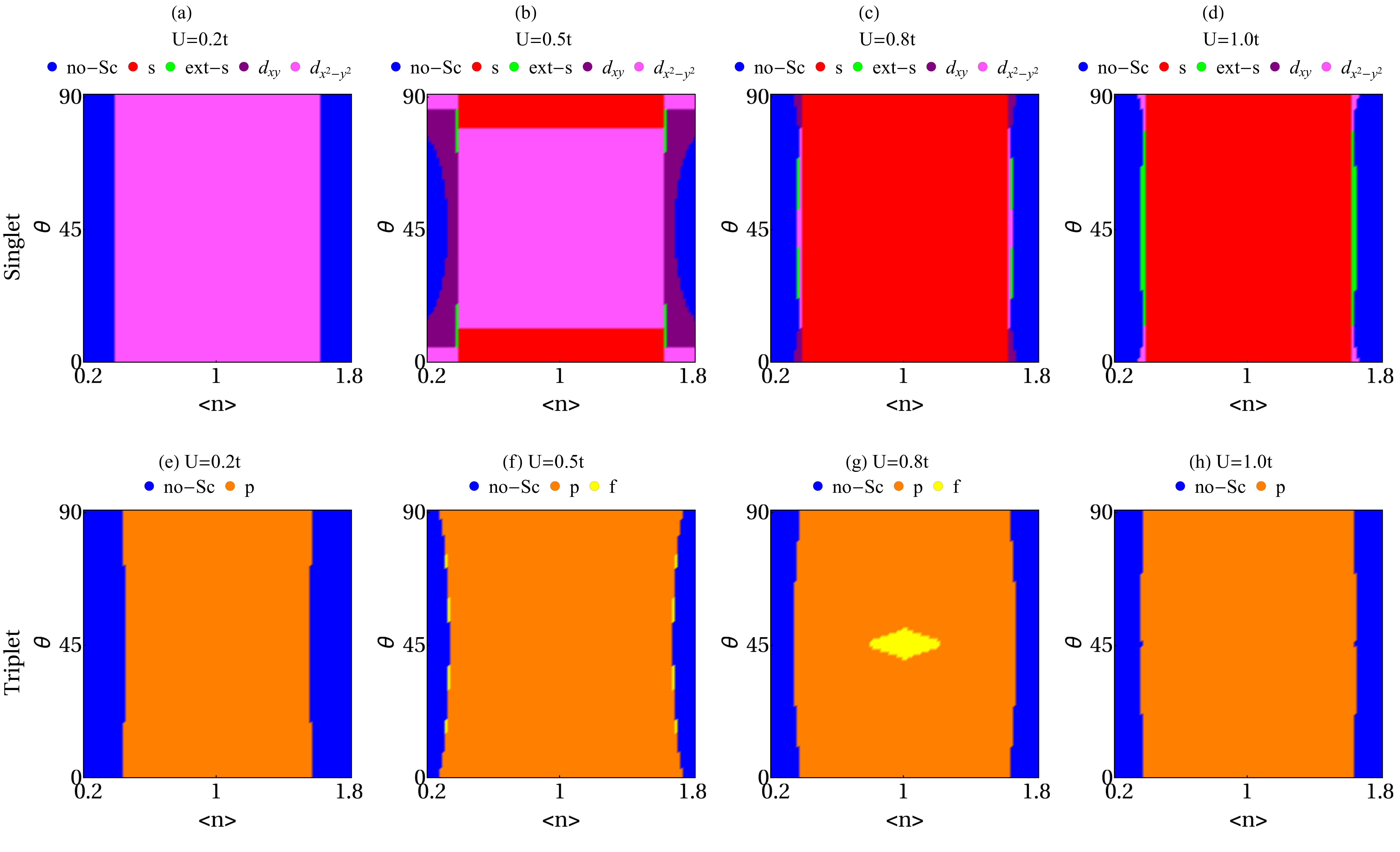} 
	\end{center}
	\vspace{-0.3cm}
	\caption{(Color online) 
		Phase diagrams derived from the dependence of the effective superconducting coupling constant $\lambda^{\rm eff}_i$ for the singlet (upper row) and triplet (lower row) channels to the level of filling and relative amplitude of Rashba and \DsH, at distinctive values of on-site Hubbard interaction.
		It is observed that for weak interactions, $d^{}_{x^2-y^2}$-wave pairing is dominant over the other symmetries in singlet channel.
		The enhance of U makes $s$-wave pairing dominant in singlet channel.
		However, in the triplet part of Cooper pairs, $p$-wave always dominates, except for a very narrow region at $U=0.8t$, near the half-filling for the same amplitudes of Rashba and \DsH, where the $f$-wave pairing is leading.
	}
	\label{Fig:Lambda_g}
\end{figure*}
%

\subsection*{Superconducting pairing analysis}
From the BCS theory, the superconducting order parameter at temperature $T$ is obtained by self-consistently solution of gap function equation 
%
\begin{equation}
\Delta^{}_{\bk\xi}=
-\frac{1}{V}\sum_{\bk'\xi'}
\Gamma^{\rm eff}_{\xi\xi'}(\bk,\bk')
\frac{\Delta^{}_{\bk'\xi'}}
{2E^{}_{\bk,\xi'}}
\tanh\frac{E^{}_{\bk,\xi'}}{2T}.
\label{Eq:BCS_gap}
\end{equation}
%
In the above equation, $\Delta^{}_{\bk\xi}$ is the superconducting gap function, 
and $E^{}_{\bk,\xi'}=\sqrt{\varepsilon^{2}_{\bk,\xi'}+\Delta^{2}_{\bk,\xi'}}$ is the corresponding  quasiparticle energy of helicity $\xi$.
Moreover, $\Gamma^{\rm eff}_{\xi\xi'}(\bk,\bk')$ denotes the effective interaction between two electrons with momenta $\bk$, $\bk'$ and helicities $\xi$, $\xi'$.
Close to the critical temperature $T_c$, one can consider  $E^{}_{\bk,\xi'}\sim |\varepsilon^{}_{\bk,\xi'}|$ and
 linearize the 
 self-consistent
 superconducting gap equation for both singlet and triplet channels
%
\bea
\bl
\Delta^{}_{\bk\xi}=
-\ln
 \Big(
 \frac{1.13\omega_c}{T_c}
 \Big)
 \sum_{\xi'}
 \int^{}_{{\rm FS}^{}_{\xi'}}
 \frac{d\bk'}{|v^{\xi'}_{\rm F}(\bk')|}
 \Gamma^{\rm eff}_{s/t}(\bk,\bk')
 \Delta^{}_{\bk'\xi'}.
\label{Eq:Linearized_gap}
\no
\el
\\
\eea
%
Here, $v^{\xi}_{\rm F}=|\nabla \varepsilon^{}_{\bk\xi}|$ is the momentum-dependent Fermi velocity for Fermi surface with helicity $\xi$,
also $\Gamma^{\rm eff}_{s}(\bk,\bk')=\Gamma^{\rm eff}_{\rm opp}(\bk,\bk')$, and $ \Gamma^{\rm eff}_{t}(\bk,\bk')=\Gamma^{\rm eff}_{\rm same}(\bk,\bk')$  correspond  to singlet and triplet channels, respectively.
\\
Converting the Eq.~(\ref{Eq:Linearized_gap}) to an eigenvalue problem, 
and
solving the gap equation reduces to find a dimensionless coefficient $\lambda^{\xi\xi'}_i$ for all possible angular momenta~$l$,  i.e.,
%
%
\bea
\bl
\lambda_i^{\xi\xi'}
\!
=
-\frac
{
\int^{}_{{\rm FS}^{}_{\xi}}
\frac{d\bk}{|v^{\xi}_{\rm F}(\bk)|}
\int^{}_{{\rm FS}^{}_{\xi'}}
\frac{d\bk'}{|v^{\xi'}_{\rm F}(\bk')|}
\eta^{}_i(\bk)
\Gamma^{\rm eff}_{s/t}(\bk,\bk')
\eta^{}_i(\bk')
}
{
\int^{}_{{\rm FS}^{}_{\xi'}}
\frac{d\bk'}{|v^{\xi'}_{\rm F}(\bk')|}
\eta^2_i(\bk')
},
\no
\el
\\
\label{Eq:Lambda}
\eea
%
where, $\lambda^{\xi\xi'}_{i}$ represents a $2\times 2$-matrix, whose  diagonal and off-diagonal elements represent the intra- and inter-band pairing amplitudes, respectively.
The effective superconducting coupling constant $\lambda^{\rm eff}_i$ for a given pairing channel~$i$ is obtained by the largest eigenvalue of the matrix~$\lambda^{\xi\xi'}_i$.
The eigenvector corresponding to the largest eigenvalue determines the symmetry of gap function.
Furthermore, the critical temperature $T_c$ assigned to $\lambda^{\rm eff}_i$ is defined as $T_c \propto \exp(-1/\lambda^{\rm eff}_i)$.
Therefore, the largest positive $\lambda^{\rm eff}_i$ generates the higher critical temperature.
Notice that in Eq.~(\ref{Eq:Lambda}), the momenta $\bk$ and $\bk'$ are restricted to the Fermi surfaces $\xi$ and $\xi'$, respectively.
Moreover, only the momentum-dependence of superconducting gap is taken into account, i.e. 
$$\Delta^{}_{\xi}(\bk)=\Delta^{}_{\xi} \; \eta(\bk).$$
 The momentum dependence of the possible superconducting pairings, $\eta^{}_i(\bk)$, at the lowest angular momenta are listed as following
%
%
%
\bea
\begin{aligned}
&
\eta^{}_{s}=1;
\hspace{2.35cm}
\eta^{}_{{\rm ext}-s}=\cos k_x+\cos k_y,
\\
&
\eta^{}_{d^{}_{xy}}=\sin k_x \sin k_y;
\hspace{0.5cm}
\eta^{}_{d^{}_{x^2-y^2}}=\cos k_x-\cos k_y,
\\
&
\eta^{}_{p}=\sin k_x;
\hspace{1.75cm}
\eta^{}_{f}=\sin k_x(\cos k_x-\cos k_y).
\hspace{.35cm}
\no
\end{aligned}
\\
\label{Eq:eta}
\eea
%

Using the above recipe,  Fig.~\ref{Fig:Lambda_g_0} represents variation of the effective coupling parameter $\lambda^{\rm eff}_i$ regarding the level of filling $\langle n \rangle$ for a system with spatial inversion symmetry (${\rm g}=0$) at $U=0.5t$.
For a wide range of filling, the $d_{x^2-y^2}$-wave pairing dominates entire the singlet channel.
However, in a small region far away from the half-filling, the singlet channel acquires $d_{xy}$-wave symmetry.
The presence of spatial inversion symmetry guarantees the negligible triplet pairing in this situation.

By the same token, Fig.~\ref{Fig:Lambda_g} illustrates the phase diagrams extracted from the values effective coupling constant $\lambda^{\rm eff}_i$ for various superconducting gap symmetries as a function of filling $\langle n \rangle$ and relative amplitudes of antisymmetric SOCs, for different values of on-site Hubbard interaction.
As shown in Fig.~\ref{Fig:Lambda_g}(a) for   the singlet channel, and  $U=0.2t$, 
the $d^{}_{x^2-y^2}$-wave pairing dominates over the system.
Interestingly, the increasing of Hubbard interaction leads to appearance of other spin-singlet pairings, especially for $U=0.5t$ [see Fig.~\ref{Fig:Lambda_g}(b)].
For a wide region centered around half-filling, near the pure Rashba (\DsH), $s$-wave pairing realizes.
However, in the regions far from the pure Rashba (\DsH), $d_{x^2-y^2}$-wave pairing dominates again.
Very far from half-filling, for many different values of Rashba and \DsH, the even-parity $d_{xy}$-wave pairing emerges. 
As they are shown in Figs.~\ref{Fig:Lambda_g}(c,~and~d),
for the cases with $U=0.8t$ and $U=1.0t$, $s$-wave pairing is almost dominant in the singlet channel and only within very narrow regions, one can observe the other angular momenta.
In addition, the dark blue region could be related to the unstable or higher order angular momentum pairings.

To complete this discussion, we have implemented our results for the triplet channel in the second row of  Fig.~\ref{Fig:Lambda_g}.
For such a case, the $p$-wave pairing is always observed, unless for a small region at $U=0.8t$, near the half filling with the same magnitudes of Rashba and \DsH, where $f$-wave pairing is established.
It should be noted that for all cases, since $\lambda^{\rm eff}_{\rm singlet}\gg \lambda^{\rm eff}_{\rm triplet}$, then the singlet pairings are the leading pairing channels.
This is why only the effect of spin fluctuations on formation of Cooper pairs are taken into account and the contributions of charge fluctuations and second neighbor hopping, which significantly strengthen the triplet pairing~\cite{Greco_Schnyder_PRL_2018,Ghadimi_PRB_2019}, are neglected.
It shows an obvious  symmetry between hole- and electron doping parts that is resulted from the particle-hole symmetry in the Hamiltonian of Eq.~(\ref{Eq:Ham}). 
%

%

\section{Even- and odd-frequency superconducting}
\begin{figure}[t]
	\begin{center}
		\hspace{-0.1cm}
		\includegraphics[width=0.98\linewidth]{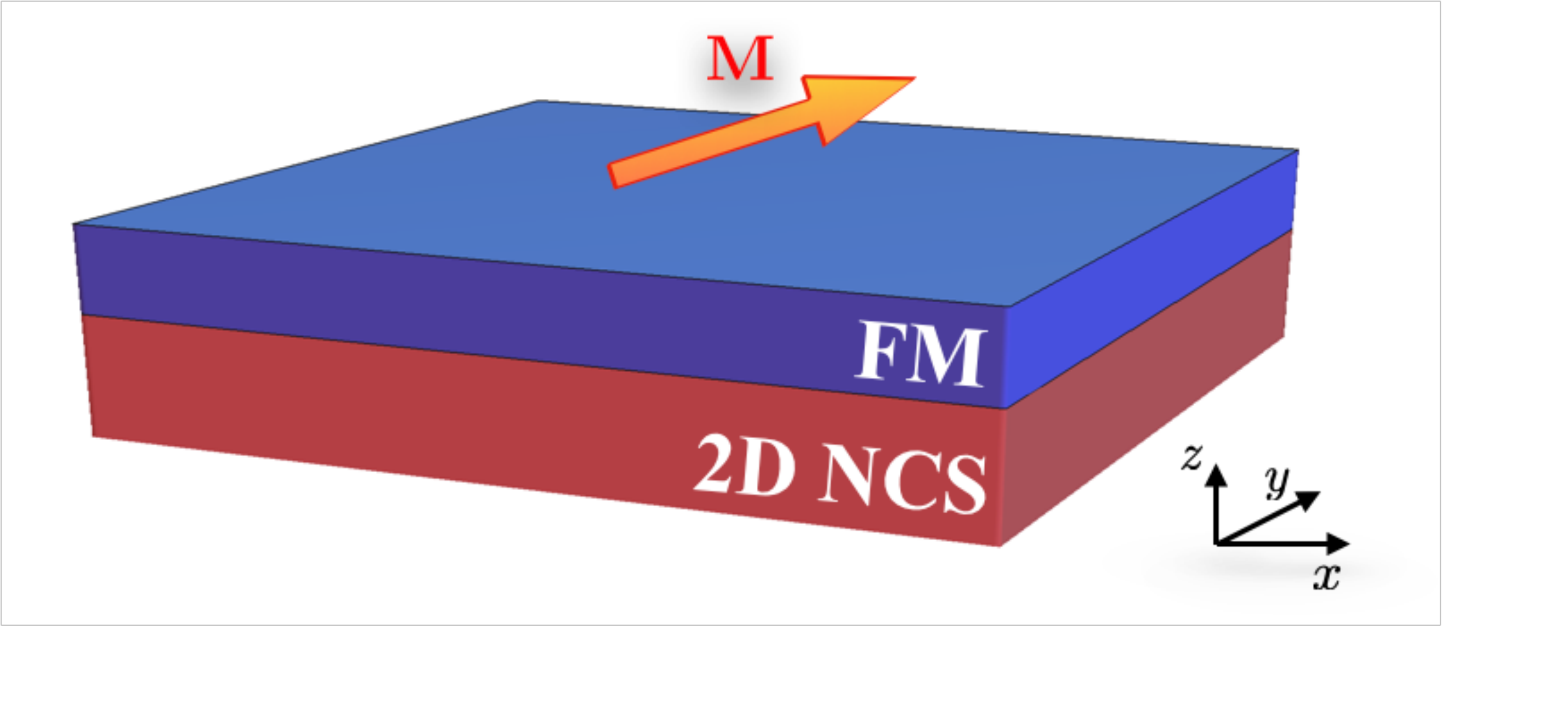} 
	\end{center}
	\vspace{-0.93cm}
	\caption{(Color online) 
		Schematic diagram of the NCS-FM heterostructure 
		considered in this work. 
		The NCS and FM occupy the $z<0$ and $z>0$
		half spaces, respectively. 
		The FM magnetization, $\bM$, points in the arbitrary direction,
		$\bM=|\bM|
		(\hat{x} \sin \theta \cos \phi 
		+\hat{y} \sin \theta \sin \phi 
		+\hat{z} \cos \theta 
		)$, 
		where $\theta$ and $\phi$ are polar and azimuthal angles, respectively.
	}
	\label{Fig:Proximity}
\end{figure}

Now we proceed to investigate the modification of Cooper pairing, resulting from the proximity with a ferromagnetic layer with magnetization vector $\bM=(M^{}_x,M^{}_y,M^{}_z)$  [see Fig.~\ref{Fig:Proximity}].
The interface is assumed to be spin-inactive, which forbids spin-flipping during the tunneling process.
 In a ferromagnet, the spins of individual atoms are coupled by the direct- or superexchange-interactions.
 The resulting net magnetic moment oriented along the easy direction leads to an effective exchange field in the ordered phase. 

 The Hamiltonian characterizing the heterostructure made by the proximity of a ferromagnet and superconductor can be written as 
 %
 \be
 \bl
 {\cal H}={\cal H}^{\rm 2D}_{\rm NCS}+{\cal H}^{\rm 2D}_{\rm FM}+{\cal H}^{}_{\rm t}.
 \el
 \ee
%
%
 %
 Here, ${\cal H}^{\rm 2D}_{\rm NCS}$ stands for the Hamiltonian of two-dimensional noncentrosymmetric supercondcutor that is described by the Bogoliubov-de Gennes (BdG) Hamiltonian, which is given by
 %
\begin{align}
\begin{aligned}
{\cal H}^{\rm 2D}_{\rm NCS}=
\frac{1}{2}\sum_{\bk}
\Psi^{\dagger}_{\bk}
{\cal H}^{\rm BdG}_{\bk}
\Psi^{}_{\bk}
,
\end{aligned}
\label{Eq:H_2D_NCS}
\end{align}
%
where 
$\Psi^{\dagger}_{\bk}=
(c^{\dagger}_{\bk\uparrow},
c^{\dagger}_{\bk\downarrow},c^{}_{\bk\uparrow},c^{}_{\bk\downarrow})$ is the Nambu space creation field operator,
 and 
${\cal H}^{\rm BdG}_{\bk}$
is defined by
\
\begin{equation}
{\cal H}^{\rm BdG}_{\bk}=
\begin{bmatrix}
\hat{h}^{}_{\bk} & \hat{\Delta}^{}_{\bk}
\\
\hat{\Delta}^{\dagger}_{\bk} & -\hat{h}^{*}_{-\bk}
\end{bmatrix}.
\label{Eq:H_BdG}
\end{equation}
%
Here the $\hat{h}^{}_{\bk}$ term 
represents the noninteracting part of Eq.~(\ref{Eq:Ham}).
Furthermore, 
$$\hat{\Delta}^{}_{\bk}={\mi}[\psi^{}_{\bk}\sigma^{}_{0}+{\bg}^{}_{\bk} \cdot \boldsymbol{\sigma}]\sigma^{}_y
$$
is the superconducting gap function in spin basis including spin-singlet $\psi^{}_{\bk}$ and spin-triplet $\bd^{}_{\bk}=(d^{x}_{\bk},d^{y}_{\bk},d^{z}_{\bk})$
pairings.
The Hamiltonian of ferromagnet is expressed as
%
\begin{align}
\begin{aligned}
{\cal H}^{\rm 2D}_{\rm FM}=
\frac{1}{2}\sum_{\bq}
\Phi^{\dagger}_{\bq}
{\cal H}^{\rm FM}_{\bq}
\Phi^{}_{\bq},
\end{aligned}
\label{Eq:H_2D_FM}
\end{align}
%
in which 
$\Phi^{\dagger}_{\bq}=
(a^{\dagger}_{\bq\uparrow},
a^{\dagger}_{\bq\downarrow},a^{}_{\bq\uparrow},a^{}_{\bq\downarrow})$ is the creation field operator of the ferromagnet with $a^{\dagger}_{\bq\sigma}$ ($a^{}_{\bq\sigma}$) as the creation (annihilation) operator of a spin-polarized electron with momentum $\bq=(q^{}_x,q^{}_y)$ and spin $\sigma$.
Moreover, ${\cal H}^{\rm FM}_{\bq}$ is
\
\begin{equation}
{\cal H}^{\rm FM}_{\bq}=
\begin{bmatrix}
\hat{h}^{\rm FM}_{\bq} & 0
\\
0 & -\hat{h}^{*\rm FM}_{-\bq}
\end{bmatrix}.
\label{Eq:H_FM_mat}
\end{equation}
%
%
In the above equation, the Hamiltonian $\hat{h}^{\rm FM}_{\bq}$ is defined~by
\begin{equation}
\hat{h}^{\rm FM}_{\bq}=
\varepsilon'_{\bq}\sigma^{}_0
-\bM\cdot \boldsymbol{\sigma},
\label{Eq:H_FM}
\end{equation}
%
where $\varepsilon'_{\bq}=\frac{\bq^2}{2m_e}-\mu_F$ is the energy dispersion of a spin-polarized electron of mass $m_e$ with chemical potential $\mu_F$.
Considering the possibility of electron tunneling between 2D-NCS and FM, the tunneling Hamiltonian is given by
\begin{align}
\begin{aligned}
{\cal H}^{}_{\rm t}=
\sum_{\bk\bq}
\Psi^{\dagger}_{\bk}
{\cal H}^{\rm t}_{\bk,\bq}
\Phi^{}_{\bq}
+
{\rm h.c.},
\end{aligned}
\label{Eq:H_tunneling}
\end{align}
%
in which 
$${\cal H}^{\rm t}_{\bk,\bq}=t'(\tau^{}_z \otimes \sigma^{}_0),$$ 
Here $t'$ is the tunneling amplitude between FM and 2D-NCS, and 
 $\tau^{}_z$ represents  the Pauli matrix in a basis consisting of the ferromagnet and superconductor.
%

%
%

%
\begin{figure}[b]
	\begin{center}
		\vspace{0.cm}
		\hspace{-0.1cm}
		\includegraphics[width=1.001 \linewidth]{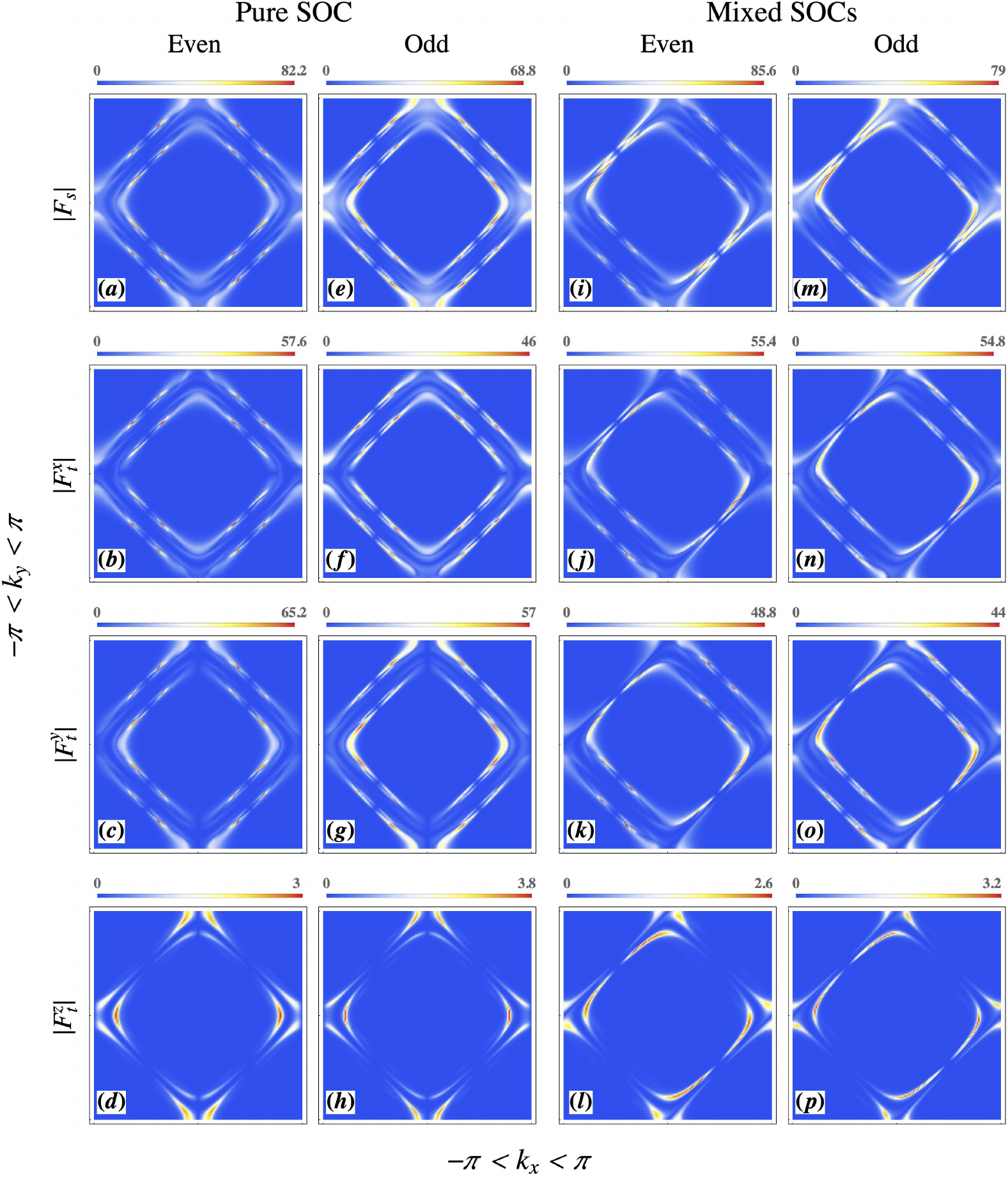} 
	\end{center}
	\vspace{-0.3cm}
	\caption{(Color online) 
		The momentum dependence of the amplitudes for even- and odd-frequency parts of modified pairing functions in a system with pure Rashba (\DsH) (left panel) and the same contributions of Rashba and \DsH~  (right panel), at $\omega=0.2t$.
		The results correspond to the filling $\langle n \rangle=0.9$
		and an exchange field with magnetization vector $\bM=(0.01t,0,0)$.
		The first row corresponds to singlet, and the second, third and forth rows represent the triplet components: $|F^{i}_t(\bk,\omega)|$; $i=x,y$, and $z$, respectively.
	}
	\label{Fig:Even_Odd_BZ}
\end{figure}
%

\subsection*{Formalism of  the Cooper pairing of 2D-noncentrosymmetric superconductor in  proximity of  a ferromagnet}
In this section we aim to investigate the modification of Cooper pairs as the result of proximity of 2D-NCS and FM with an arbitrary magnetization $\bM$.
The unperturbed Green's function of 2D-NCS can be written~as
\begin{align}
\check{\cal G}^{0}_{\rm NCS}(\bk,{\mi}\omega_n)=
\begin{bmatrix}
\hat{\cal G}^{0}_{\rm NCS}(\bk,{\mi}\omega_n)
&
\hat{\cal F}^{0}_{\rm NCS}(\bk,{\mi}\omega_n)
\\
\hat{\cal F}^{0\dagger}_{\rm NCS}(\bk,{\mi}\omega_n)
&
-\hat{\cal G}^{0\intercal}_{\rm NCS}(-\bk,-{\mi}\omega_n)
\end{bmatrix},
\label{Eq:Green_NCS}
\end{align}
%
where 
$\hat{\cal G}^{0}_{\rm NCS}(\bk,{\mi}\omega_n)$ and $\hat{\cal F}^{0}_{\rm NCS}(\bk,{\mi}\omega_n)$ are unperturbed normal and anomalous Green's functions of 2D-NCS~\cite{Frigeri_Sigrist_PRL_2004,Akbari_EPL_2013,*Akbari:2013aa}, which are defined~by
\begin{align}
\begin{aligned}
\hat{\cal G}^{0}_{\rm NCS}(\bk,{\mi}\omega_n)
&=
{\cal G}^{0}_{+}(\bk,{\mi}\omega_n)\sigma^{}_0
+
(\hat{\bg}^{}_{\bk} \cdot \boldsymbol{\sigma})
{\cal G}^{0}_{-}(\bk,{\mi}\omega_n),
\\
\hat{\cal F}^{0}_{\rm NCS}(\bk,{\mi}\omega_n)
&=
\!
\Big[
{\cal F}^{0}_{+}(\bk,{\mi}\omega_n)\sigma^{}_0
+
(\hat{\bg}^{}_{\bk} \cdot \boldsymbol{\sigma})
{\cal F}^{0}_{-}(\bk,{\mi}\omega_n)
\Big]
\hat{\Delta}^{}_{\bk},
\end{aligned}
\label{Eq:Normal_Anomalous_G_NCS}
\end{align}
%
with
\begin{align}
\begin{aligned}
{\cal G}^{0}_{\xi}(\bk,{\mi}\omega_n)
&=
-\frac{1}{2}
\sum_{\xi}
\xi
\frac{
	{\mi}\omega_n+\varepsilon^{}_{\bk\xi}
}
{
\omega^2_n+\varepsilon^{2}_{\bk\xi}+\Delta^{2}_{\bk\xi}
}
,
\\
{\cal F}^{0}_{\xi}(\bk,{\mi}\omega_n)
&=
\frac{1}{2}
\sum_{\xi}
\xi
\frac{
	1
}
{
	\omega^2_n+\varepsilon^{2}_{\bk\xi}+\Delta^{2}_{\bk\xi}
}.
\end{aligned}
\label{Eq:G_F_NCS_pm}
\end{align}
%
Here 
$$\Delta^{}_{\bk\xi}=\psi^{}_{\bk}+\xi|\bd^{}_{\bk}|$$
 is the magnitude of the superconducting gap for FS with helicity $\xi$.
The Green's function of FM is written as
\begin{align}
\check{\cal G}^{0}_{\rm FM}(\bq,{\mi}\omega_n)=
\begin{bmatrix}
\hat{\cal G}^{0}_{\rm FM}(\bq,{\mi}\omega_n)
&
0
\\
0
&
-\hat{\cal G}^{0\intercal}_{\rm FM}(-\bq,-{\mi}\omega_n)
\end{bmatrix},
\label{Eq:Green_FM}
\end{align}
%
where
\begin{align}
\begin{aligned}
\hat{\cal G}^{0}_{\rm FM}(\bq,{\mi}\omega_n)
&=
\Big[
	{\mi}\omega_n \sigma^{}_0-\hat{h}^{\rm FM}_{\bq}
\Big]^{-1}_{}
\\
&=
\frac
{
	({\mi}\omega_n-\varepsilon'_{\bq})\sigma^{}_0
	-
	\bM \cdot \boldsymbol{\sigma}}
{
({\mi}\omega_n-\varepsilon'_{\bq})^2_{}-|\bM|^2_{}
}.
\end{aligned}
\label{Eq:G_FM}
\end{align}
%

The total Green's function of the 2D-NCS can be read using the following Dyson equation
\bea
\check{\cal G}^{}_{\rm NCS}(\bk,{\mi}\omega_n)
=
\check{\cal G}^{0}_{\rm NCS}(\bk,{\mi}\omega_n)
\Big[
\mathbbm{1}
+
\check{\Sigma}(\bk,{\mi}\omega_n)
\check{\cal G}^{}_{\rm NCS}(\bk,{\mi}\omega_n)
\Big].
\label{Eq:Dyson}
\no
\\
\eea
%
Here the self-energy term $\check{\Sigma}(\bk,{\mi}\omega_n)$, carrying the effect of the magnetic proximity, is given by
\begin{equation}
\check{\Sigma}(\bk,{\mi}\omega_n)
=\sum_{\bq}
{\cal H}^{\rm t}_{\bq}
\;
\check{\cal G}^{0}_{\rm FM}(\bq,{\mi}\omega_n)
\;
{\cal H}^{\rm t \dagger}_{\bq}.
\end{equation}
%
It is worthwhile to be mentioned that momentum conservation forces the tunneling phenomena between NCS and FM to occur only for the electrons with the same momenta ($\bk=\bq$)~\cite{Triola_PRB_2014}.  
Within the weak coupling regime, the FM proximity minimally affects the electronic band structure of 2D-NCS.
Thus, up to the second order perturbation, total Green's function of 2D-NCS is described by
\bea
\check{\cal G}^{}_{\rm NCS}(\bk,{\mi}\omega_n)
\approx
\check{\cal G}^{0}_{\rm NCS}(\bk,{\mi}\omega_n)
\Big[
\mathbbm{1}
+
\check{\Sigma}(\bk,{\mi}\omega_n)
\check{\cal G}^{0}_{\rm NCS}(\bk,{\mi}\omega_n)
\Big].
\label{Eq:Green_2nd_Perturbation}
\no
\\
\eea
%
Therefore, it is possible to rewrite the anomalous Green's function of the modified 2D-NCS as
\begin{equation}
\hat{{\cal F}}(\bk,{\mi}\omega_n)=
\Big[
F^{}_s(\bk,{\mi}\omega_n)\sigma^{}_0
+
{\bF}^{}_t(\bk,{\mi}\omega_n)
\cdot
\bd^{}_{\bk}
\Big]
{\mi}\sigma^{}_y,
\label{Eq:Anomalous}
\end{equation}
%
where the singlet and spatial components of triplet pairings are defined by
\begin{widetext}
\begin{align}
\begin{aligned}
&
F^{}_{s}(\bk,{\mi}\omega_n)
=
\frac{1}{2}
\Big[
F^{}_{\uparrow\downarrow}(\bk,{\mi}\omega_n)
-
F^{}_{\downarrow\uparrow}(\bk,{\mi}\omega_n)
\Big]
;
\hspace{1.25cm}
F^{x}_{t}(\bk,{\mi}\omega_n)
=
\frac{1}{2}
\Big[
F^{}_{\downarrow\downarrow}(\bk,{\mi}\omega_n)
-
F^{}_{\uparrow\uparrow}(\bk,{\mi}\omega_n)
\Big],
\\
&
F^{y}_{t}(\bk,{\mi}\omega_n)
=
\frac{-{\mi}}{2}
\Big[
F^{}_{\uparrow\uparrow}(\bk,{\mi}\omega_n)
+
F^{}_{\downarrow\downarrow}(\bk,{\mi}\omega_n)
\Big]
;
\hspace{1cm}
F^{z}_{t}(\bk,{\mi}\omega_n)
=
\frac{1}{2}
\Big[
F^{}_{\uparrow\downarrow}(\bk,{\mi}\omega_n)
+
F^{}_{\downarrow\uparrow}(\bk,{\mi}\omega_n)
\Big].
\end{aligned}
\label{Eq:Anomal_Components}
\end{align}
%
Based of Eq.(~\ref{Eq:Anomal_Components}), the final forms of modified singlet and triplet superconductor in 2D-NCS are given by
%
\begin{align}
\begin{aligned}
F^{}_{s}(\bk,{\mi}\omega_n)
=&
{\cal F}^{0}_{-}(\bk,{\mi}\omega_n)
\Big[
{\cal I}(\bk)
+
t'^2
\Big\lbrace
{\cal G}^{0}_{-}(-\bk,-{\mi}\omega_n)
{\cal J}(\bk,{\mi}\omega_n)
+
{\cal G}^{0}_{-}(\bk,{\mi}\omega_n)
{\cal K}(\bk,{\mi}\omega_n)
\Big\rbrace
\Big],
\\
F^{x}_{t}(\bk,{\mi}\omega_n)
=&
{\cal F}^{0}_{+}(\bk,{\mi}\omega_n)
\Big[
\Big\lbrace
1+
t'^2
\Big\lbrace
{\cal G}^{0}_{+}(\bk,{\mi}\omega_n)
{\cal L}(\bk,{\mi}\omega_n)
+
{\cal G}^{0}_{+}(-\bk,-{\mi}\omega_n)
{\cal L'}(\bk,{\mi}\omega_n)
\Big\rbrace
\Big\rbrace
d^{}_{x}(\bk)
\\
&-{\mi}t'^2_{} M^{}_z
\Big\lbrace
{\cal G}^{0}_{+}(\bk,{\mi}\omega_n)
{\cal M}(\bk,{\mi}\omega_n)
+
{\cal G}^{0}_{+}(-\bk,{-\mi}\omega_n)
{\cal M'}(\bk,{\mi}\omega_n)
\Big\rbrace
\Big],
\\
F^{y}_{t}(\bk,{\mi}\omega_n)
=&
-{\cal F}^{0}_{+}(\bk,{\mi}\omega_n)
\Big[
\Big\lbrace
1+
t'^2
\Big\lbrace
{\cal G}^{0}_{+}(\bk,{\mi}\omega_n)
{\cal L}(\bk,{\mi}\omega_n)
+
{\cal G}^{0}_{+}(-\bk,-{\mi}\omega_n)
{\cal L'}(\bk,{\mi}\omega_n)
\Big\rbrace
\Big\rbrace
d^{}_{y}(\bk)
\\
&+{\mi}t'^2_{} M^{}_z
\Big\lbrace
{\cal G}^{0}_{+}(\bk,{\mi}\omega_n)
{\cal M}(\bk,{\mi}\omega_n)
+
{\cal G}^{0}_{+}(-\bk,{-\mi}\omega_n)
{\cal M'}(\bk,{\mi}\omega_n)
\Big\rbrace
\Big],
\\
F^{z}_{t}(\bk,{\mi}\omega_n)
=&
{\cal F}^{0}_{-}(\bk,{\mi}\omega_n)
\Big[
{\cal I'}(\bk)
+
t'^2
\Big\lbrace
{\cal G}^{0}_{-}(-\bk,-{\mi}\omega_n)
{\cal J'}(\bk,{\mi}\omega_n)
+
{\cal G}^{0}_{-}(\bk,{\mi}\omega_n)
{\cal K'}(\bk,{\mi}\omega_n)
\Big\rbrace
\Big].
\end{aligned}
\label{Eq:Singlet_Triplet_components}
\end{align}
%
%
%
This analytically indicates that all components of the modified anomalous Green's function can be expressed as a combination of both even- and odd-functions in terms of frequency, ${\mi}\omega_n$.
The functions ${\cal I}(\bk)$, ${\cal I'}(\bk)$ in Eq.(~\ref{Eq:Singlet_Triplet_components}) do not depend on frequency and are defined as
\begin{align}
\begin{aligned}
{\cal I}(\bk)&=
{\rm g}^{x}_{\bk} \psi^{}_{\bk} 
-{\mi} {\rm g}^{y}_{\bk} d^{z}_{\bk};
\hspace{1cm}
{\cal I'}(\bk)&=
{\rm g}^{x}_{\bk} d^{z}_{\bk}
-{\mi} {\rm g}^{y}_{\bk} \psi^{}_{\bk}.
\end{aligned}
\label{Eq:I_Ip}
\end{align}
%
Whereas the frequency dependent functions ${\cal J}(\bk,{\mi}\omega_n)$, ${\cal J'}(\bk,{\mi}\omega_n)$, ${\cal L}(\bk,{\mi}\omega_n)$,  ${\cal L'}(\bk,{\mi}\omega_n)$, ${\cal M}(\bk,{\mi}\omega_n)$, and  ${\cal M'}(\bk,{\mi}\omega_n)$ are given by
\begin{align}
\begin{aligned}
{\cal J}(\bk,{\mi}\omega_n)&=
\frac{M^{}_x \psi^{}_{\bk}+{\mi}M^{}_y d^{z}_{\bk}}
{({\mi}\omega_n+\varepsilon'_{\bk})^2_{}-|\bM|^2_{}};
\hspace{1cm}
{\cal J'}(\bk,{\mi}\omega_n)&=
\frac{
	M^{}_x d^{z}_{\bk}+{\mi}M^{}_y \psi^{}_{\bk}
}
{
	({\mi}\omega_n+\varepsilon'_{\bk})^2_{}-|\bM|^2_{}
},
\\
{\cal L}(\bk,{\mi}\omega_n)&=
\frac{
	{\mi}\omega_n-\varepsilon'_{\bk}	
}
{({\mi}\omega_n-\varepsilon'_{\bk})^2_{}-|\bM|^2_{}};
\hspace{1cm}
{\cal L'}(\bk,{\mi}\omega_n)&=
\frac{
	{\mi}\omega_n+\varepsilon'_{\bk}	
}
{({\mi}\omega_n+\varepsilon'_{\bk})^2_{}-|\bM|^2_{}},
\\
{\cal M}(\bk,{\mi}\omega_n)&=
\frac{
	1	
}
{({\mi}\omega_n-\varepsilon'_{\bk})^2_{}-|\bM|^2_{}};
\hspace{1cm}
{\cal M'}(\bk,{\mi}\omega_n)&=
\frac{
	1
}
{({\mi}\omega_n+\varepsilon'_{\bk})^2_{}-|\bM|^2_{}},
\end{aligned}
\label{Eq:J_Jp_L_Lp}
\end{align}
%
Finally, the functions  ${\cal K}(\bk,{\mi}\omega_n)$, ${\cal K'}(\bk,{\mi}\omega_n)$ in Eq.~(\ref{Eq:Singlet_Triplet_components}) are defined as follow
%
%
\begin{align}
\begin{aligned}
{\cal K}(\bk,{\mi}\omega_n)&=
\frac
{
	2{\rm g}^{x}_{\bk} {\rm g}^{y}_{\bk} 
	(
	{\mi} M^{}_x d^{z}_{\bk}  + M^{}_y \psi^{}_{\bk}  
	)
	-
	(
	{\rm g}^{x2}_{\bk}-{\rm g}^{y2}_{\bk}
	)
	(
	M^{}_x \psi^{}_{\bk}+{\mi}M^{}_y d^{z}_{\bk}
	)
}
{
({\mi}\omega_n-\varepsilon'_{\bk})^2_{}-|\bM|^2_{}
},
\\
{\cal K'}(\bk,{\mi}\omega_n)&=
\frac
{
	2{\rm g}^{x}_{\bk} {\rm g}^{y}_{\bk} 
	(
	{\mi}  M^{}_x  \psi^{}_{\bk} +M^{}_y d^{z}_{\bk}
	)
	-
	(
	{\rm g}^{x2}_{\bk}-{\rm g}^{y2}_{\bk}
	)
	(
	M^{}_x d^{z}_{\bk}-{\mi}M^{}_y \psi^{}_{\bk} 
	)
}
{
	({\mi}\omega_n-\varepsilon'_{\bk})^2_{}-|\bM|^2_{}
}.
\end{aligned}
\label{Eq:K_Kp}
\end{align}
%
\end{widetext}
%
%

%
\begin{figure}[t]
	\begin{center}
		\vspace{0.cm}
		\hspace{-0.1cm}
		\includegraphics[width=1 \linewidth]{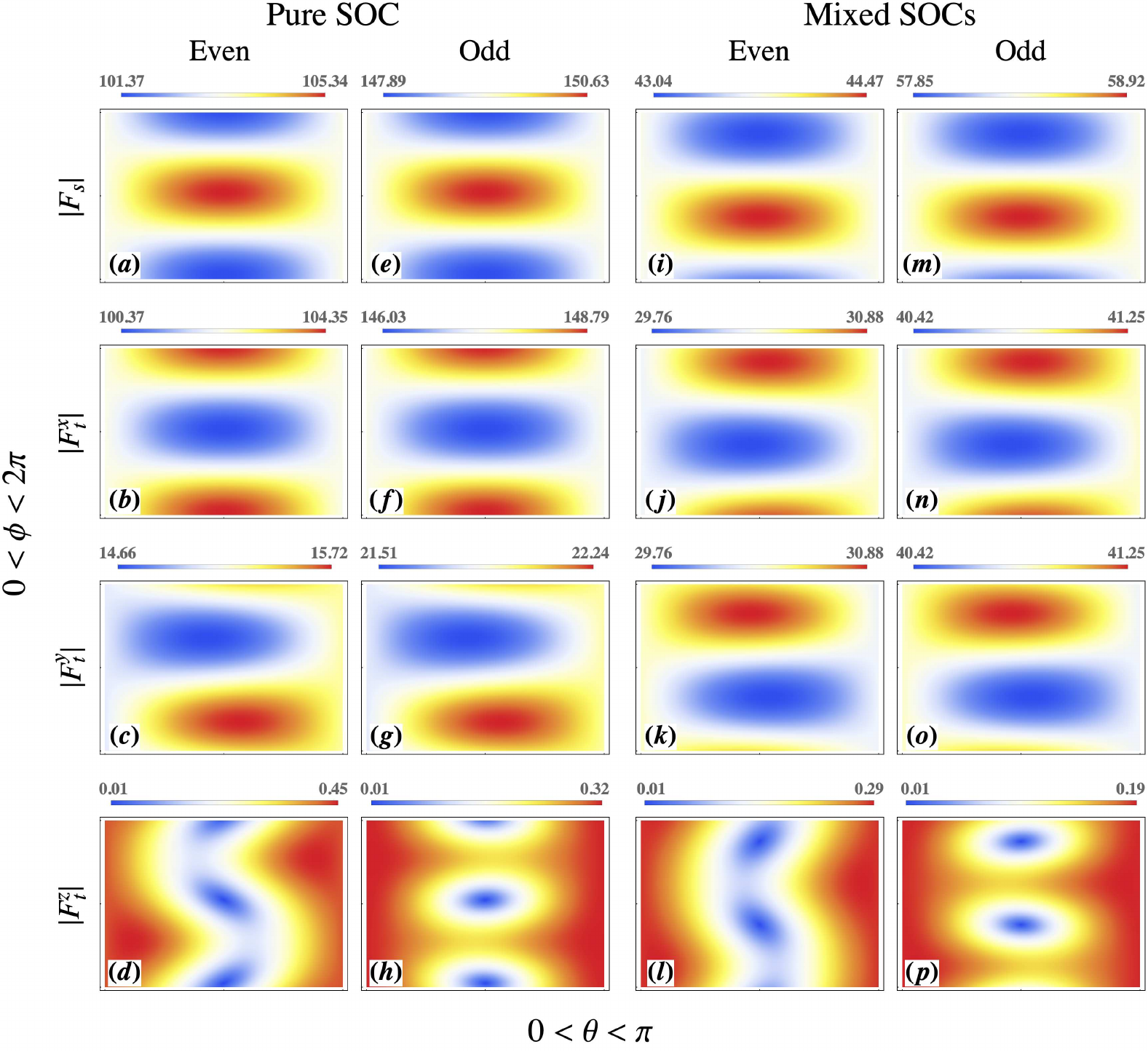} 
	\end{center}
	\vspace{-0.3cm}
	\caption{(Color online) 
		Magnitude of singlet $|F^{}_s(\bk,\omega)|$ and triplet $|F^{i}_t(\bk,\omega)|$ ($i=x,y,z$) components of anomalous Green function with respect to the direction of magnetization vector: $\theta$, and $\phi$, in 
		 the spherical coordinates:  ($|\bM|$, $\theta$, $\phi$) 
		 with magnetization amplitude  $|\bM|=0.01t$.
				 Left panel shows the results for the system with pure Rashba (\DsH), and the right panel represents them for the same mixture of Rashba and \DsH.
			Here and in the following figures, we set the momenta  at $\bk=(0.13,-2.15)$,  and the energy at $\omega=0.2t$.
	}
	\label{Fig:Theta_Phai}
\end{figure}
%
%
\begin{figure}[t]
	\begin{center}
		\vspace{0.cm}
		\hspace{-0.1cm}
		\includegraphics[width=1 \linewidth]{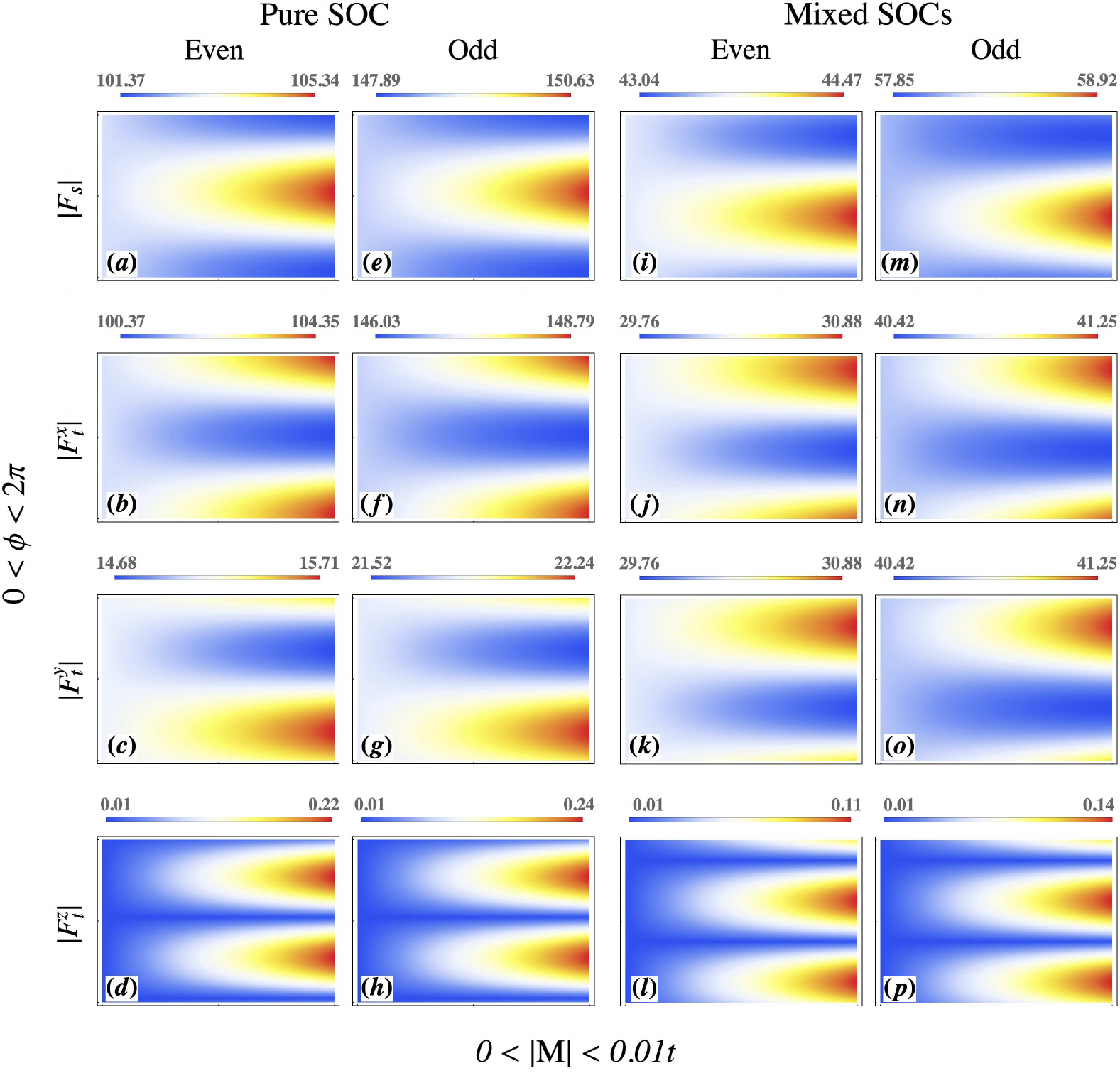} 
	\end{center}
	\vspace{-0.3cm}
	\caption{(Color online) 
		Magnitude of singlet $|F^{}_s(\bk,\omega)|$ and triplet $|F^{i}_t(\bk,\omega)|$ ($i=x,y,z$) components  of anomalous Green function, for the case of the
		magnetization vector $\bM$ aligned  in 
		``$xy$-plane",  versus the magnitude $|M|$ and  polar angle $\phi$.
		 Left panel shows the results for the system with pure Rashba (\DsH), and the right panel represents them for the same mixture of Rashba and \DsH.
		 The momenta and the energy are the same as Fig.~\ref{Fig:Theta_Phai}.
	}
	\label{Fig:J_Phai}
\end{figure}
%
%

%
\begin{figure}[t]
	\begin{center}
		\vspace{0.cm}
		\hspace{-0.1cm}
		\includegraphics[width=1 \linewidth]{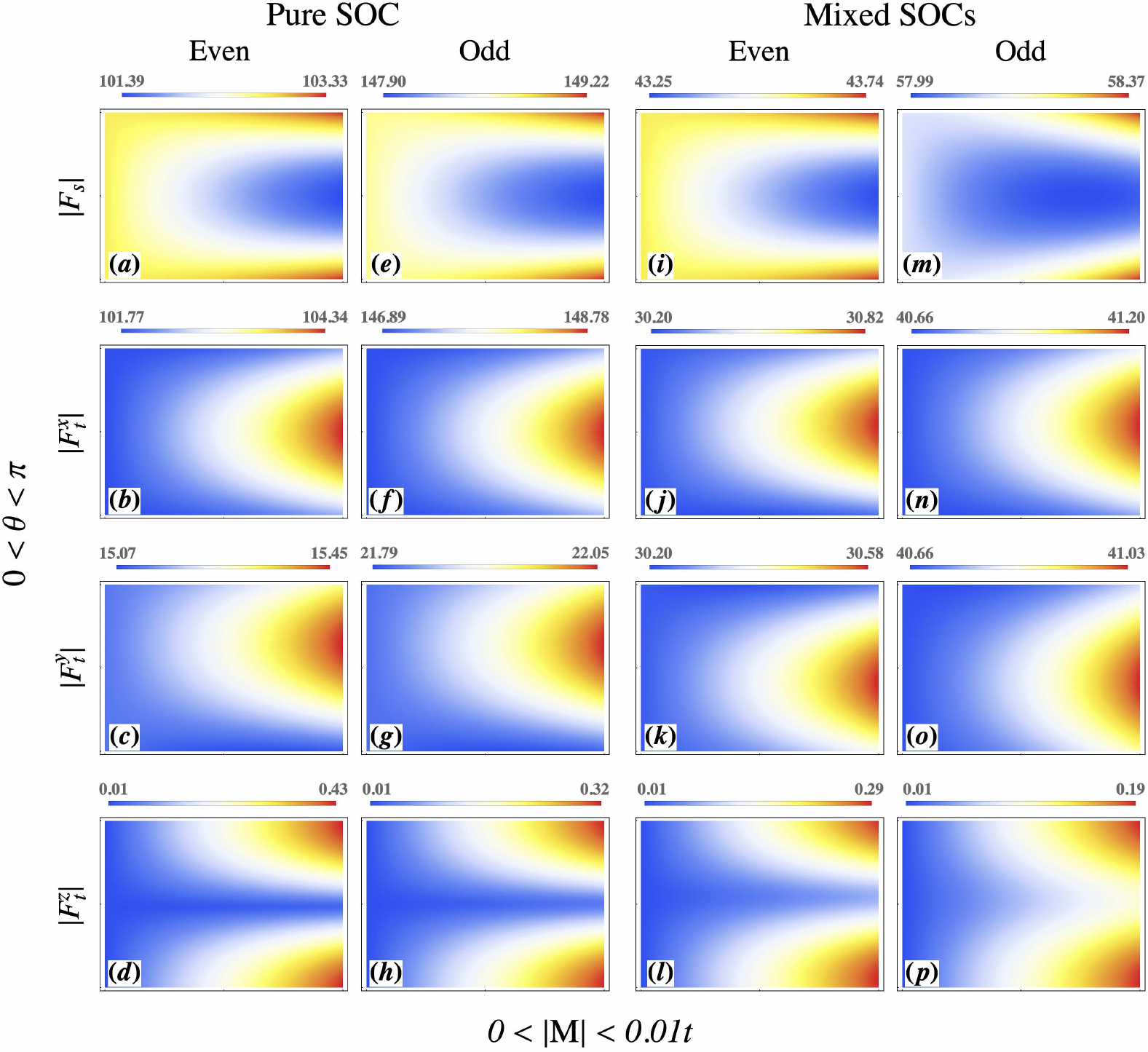} 
	\end{center}
	\vspace{-0.3cm}
	\caption{(Color online) 
		Same as Fig.~\ref{Fig:J_Phai} but for the magnetization vector $\bM$ aligned  in 
		``$xz$-plane",  with respect to the magnitude $|M|$ and azimuthal angle $\theta$.
	}
	\label{Fig:J_Theta}
\end{figure}
%
%
\begin{figure}[t]
	\begin{center}
		\vspace{0.cm}
		\hspace{-0.1cm}
		\includegraphics[width=1 \linewidth]{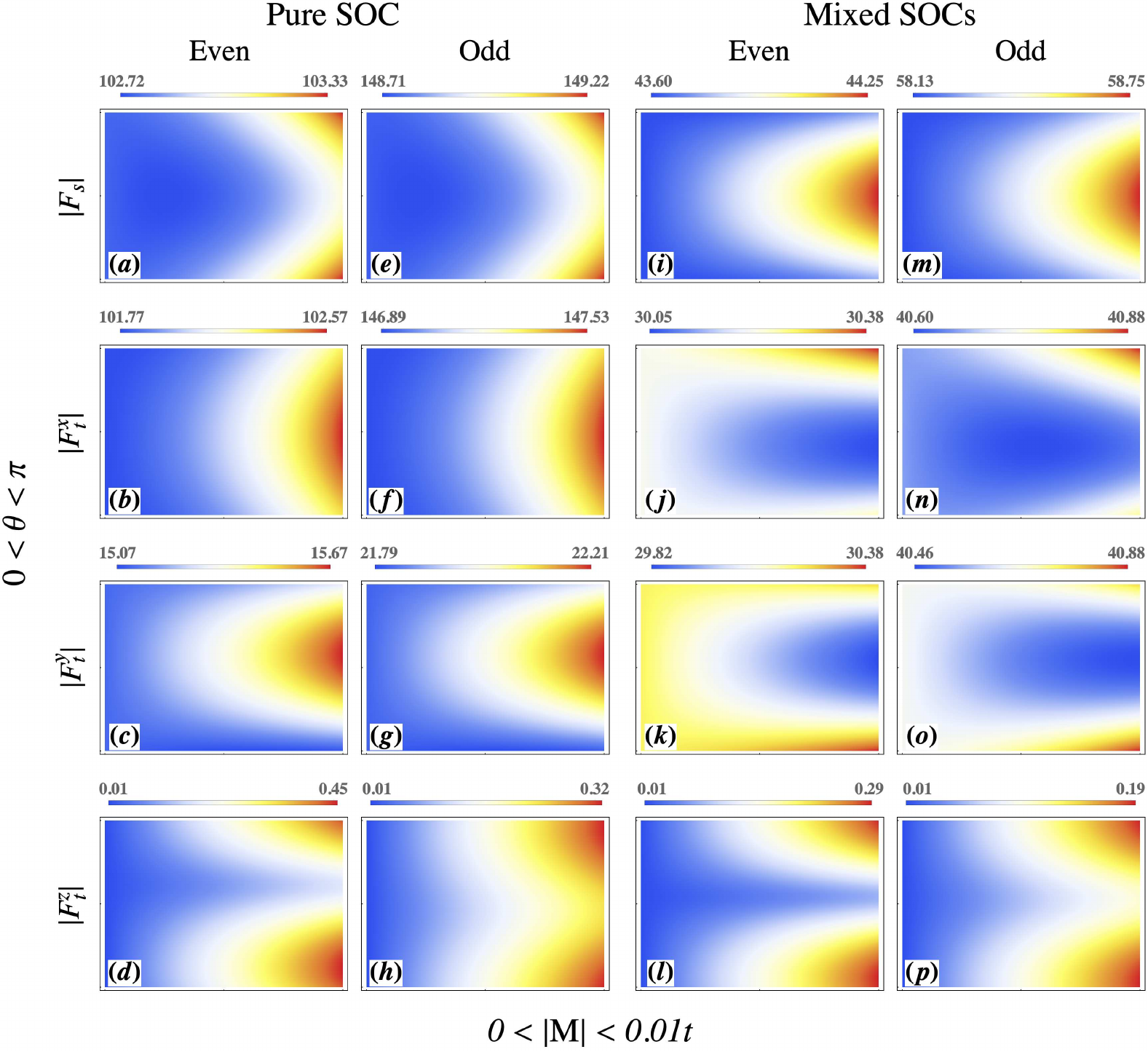} 
	\end{center}
	\vspace{-0.3cm}
	\caption{(Color online) 
		Same as Fig.~\ref{Fig:J_Phai} but for the magnetization vector $\bM$ aligned  in 
		``$yz$-plane",  with respect to the magnitude $|M|$ and azimuthal angle $\theta$.
	}
	\label{Fig:J_Theta_YZ}
\end{figure}
%

Using the mentioned equations, Fig.~\ref{Fig:Even_Odd_BZ} portrays the amplitude of modified anomalous Green's functions within BZ in the presence pure Rashba/\DsH~ (left panel), and in the presence of  the same contributions of Rashba and \DsH~  (right panel), at energy $\omega=0.2t$.

The type of superconducting gap function excluding NCS-FM proximity is obtained as $d^{}_{x^2-y^2}+p$-wave.
It is obviously observed that as the result of coupling between 2D-NCS and FM, the even- and odd-frequency pairings blends together.
One should note that in a 2D-NCS the condition $\bd^{}_{\bk}||\bg^{}_{\bk}$ should be satisfied to survive triplet pairs.
Since the antisymmetric SOC $\bg$-vector has only $x$- and $y$-components, thus the triplet $\bd$-vector has only in-plane elements.
 Figs.~\ref{Fig:Even_Odd_BZ}(d,~h,~l~and~p) illustrate that the NCS-FM proximity induces out-of-plane triplet component ($F^z_t$) into the NCS.
This is resulted from amplifying the out-of-plane ferromagnetic fluctuations due to the proximity effect.
The signature of Fermi surface is obviously detectable in both even and odd parts of the modified anomalous Green's function.
We find that the angular momentum of initial singlet and triplet pairings do not have a meaningful impact on the nature of modified even- and odd-frequency superconductivity.
\\

To see the spherical and magnetic orientation dependency of anomalous Green's functions, we sketch in the subplots of Fig.~\ref{Fig:Theta_Phai},  the impact of FM magnetization vector direction on the magnitude of even- and odd-frequency parts of the modified anomalous Green's function at an example point in the BZ and energy space with coordinates of  $\bk=(0.13,-2.15)$
and $\omega=0.2t$.	
It is obvious that coexistence of Rashba and \DsH~ considerably alters the magnitudes of $|F^{}_{s}(\bk,{\mi}\omega_n)|$ and $|F^{x,y,z}_{t}(\bk,{\mi}\omega_n)|$, for this  point.
Besides, for an in-plane magnetization vector in a system with pure Rashba (\DsH), $|F^{}_{s}(\bk,{\mi}\omega_n)|$ has almost its maximum value, however, $|F^{x,y}_{t}(\bk,{\mi}\omega_n)|$ are in their minima.
This figure clearly shows the existence of out-of-plane triplet pairing for non-zero magnetization and its variations in terms of direction of magnetization vector and alternation of antisymmetric SOCs. 
One can easily find that mixing of Rashba and \DsH~  thoroughly shifts these results.
\\

To complete our investigations, we display the variations of singlet and triplet channels of even- and odd-frequency parts of anomalous Green's function (absolute values)   in Figs.~(\ref{Fig:J_Phai},~\ref{Fig:J_Theta},~and~\ref{Fig:J_Theta_YZ}),    with respect to the magnitude of magnetization vector in the $xy$-  $xz$-, and $yz$-planes, respectively.

In $xy$-plane (Fig.~\ref{Fig:J_Phai}), increasing the magnitude of magnetization vector leads to raise the even and odd parts of modified singlet anomalous Green's function at $\phi=\pi$,  for the pure SOC.
However, this maximum slightly  shifts from this  optimum azimuthal angle  in the  case of
mixing SOCs.
Moreover, although  the $x$-component of modified triplet anomalous Green's function shows its maxima at $\phi\sim 0~$or$~\phi\sim 2\pi$ in the case of pure SOC,
 the $y$- and $z$-components behave totally different; the maximum in $y$ shifts a bit from the mentioned angles and the $z$-component shows two maxima at  $\phi\sim \pi/2~$and$~\phi\sim 3\pi/2$. 
In all cases, manipulating the relative strengths of Rashba and \DsH~  
considerably change the behaviors of $|F^{}_{s}(\bk,{\mi}\omega_n)|$ and $|F^{x,y,z}_{t}(\bk,{\mi}\omega_n)|$.

In $xz$-plane (see Fig.~\ref{Fig:J_Theta}), the enhance of  the magnetization amplitude may result in increase or decrease of the magnitude of $|F_s|$, with respect to the value of polar angle $\theta$.
For polar angles near $\theta=0$ or $\theta=\pi$, raising the magnetization enlarges $|F_s|$,
but, for $\theta\sim \pi/2$, the behavior is reversed. 
Finally, for all components of triplet pairings, the increment of magnetization vector amplifies the magnitude of modified anomalous Green's function.
At the end we should mention that the results of $xz$-plane, comprehensively, hold for the singlet and triplet parts of anomalous Green's function in $yz$-plane (see Fig.~\ref{Fig:J_Theta_YZ}).
%
%

\section{Conclusion}
 We present a study of superconducting instabilities and gap structures in a 2D noncentrosymmetric square lattice with on-site Hubbard interactions.
 We examine the influence of two important discrete symmetries, i.e.  time-reversal and inversion symmetries, on the parity and frequency characteristics of Cooper pairs.
 We analyze the consequences of Rashba and \DsH~ SOCs coincidence on the pairing symmetry at different levels of hole- and electron-doping within a spin-fluctuation-mediated scenario.
 Using RPA approach, we observe pure $d$-wave pairing at the presence of inversion symmetry, which is consistent with earlier studies~\cite{Scalapino_PRB_1986,Romer_Eremin_PRB_2015}.
In the absence of inversion symmetry, we find a dominant $d^{}_{x^2-y^2}+p$-wave pairing at low levels of interaction.
An enhancement in the strength of Hubbard term dominates $s+p$-wave pairing over a wide region of filling and relative amplitudes of Rashba and \DsH.
However, the phase diagrams of superconducting pairings (Fig.~\ref{Fig:Lambda_g}) reveal the emergence of other symmetries such as $d^{}_{xy}+p$-wave, $s^*_{}+p$-wave, and $s+f$-wave pairings in narrow regions of phase diagrams.
Particularly, in the vicinity of half-filling at $U=0.8t$, close to van-Hove singularity, for the nearly same magnitudes of Rashba and \DsH, the leading triplet pairing solution has $f$-wave symmetry, which has been originated from ferromagnetic fluctuations.
Moreover, we investigate the realization of odd- and even- frequency superconductivity at  a ferromagnet proximity in a 2D NCS.
Using the structure of anomalous Green's function up to the second order perturbation theory, we show  the admixture of both odd- and even-frequency components, resulted from the broken time-reversal symmetry.
We approach to a general expression for the pairing amplitudes of the anomalous Green's function of a heterostructure including a 2D-NCS and a FM with an arbitrary magnetization. 
Finally, we show that our theory of cooperative Rashba and \DsH~  is capable to manipulate the pairing symmetry of Cooper pairs for both parity and frequency point of views.
%

\section*{ACKNOWLEDGMENTS}
 We are grateful to I.~Eremin, P. Thalmeier  and Y.~Bang for fruitful discussions.
 This work  is supported  through National Research Foundation (NRF) funded by the Ministry of Science of Korea (Grants  No. 2017R1D1A1B03033465, \& No. 2019R1H1A2039733).
M.\ B. acknowledges the receipt of the grant No. AF-03/18-01 from Abdus Salam International Center for Theoretical Physics, Trieste, Italy.
A.A. acknowledges the National Foundation of Korea  funded by the Ministry of Science, ICT and Future Planning (No. 2016K1A4A4A01922028).

\bibliography{References}
\end{document}